\documentclass[aip,pop,preprint,floatfix,a4paper]{revtex4-1}
\usepackage{graphicx}
\usepackage{xcolor}
\usepackage{amsmath}
\usepackage[ruled,linesnumbered,vlined]{algorithm2e}
\begin{document}

\title{Extension of a multi-region free-surface MHD solver beyond the inductionless approximation}

\author{Min Ki Jung}
\affiliation{Department of Nuclear Engineering, Seoul National University, Seoul, 08826, South Korea}
\author{Brian Wynne}
\affiliation{Department of Mechanical and Aerospace Engineering, Princeton University, Princeton, New Jersey, 08540, USA}
\author{Francisco Saenz}
\affiliation{Department of Mechanical and Aerospace Engineering, Princeton University, Princeton, New Jersey, 08540, USA}
\author{Yufan Xu}
\affiliation{Princeton Plasma Physics Laboratory, Princeton, New Jersey, 08540, USA}
\author{Jabir Al-Salami}
\affiliation{Research Institute for Applied Mechanics, Kyushu University, Kasuga, Fukuoka, 816-8580, Japan}
\author{Yong-Su Na}
\email[Author to whom correspondence should be addressed: ]{ysna@snu.ac.kr}
\affiliation{Department of Nuclear Engineering, Seoul National University, Seoul, 08826, South Korea}
\author{Egemen Kolemen}
\email[Author to whom correspondence should be addressed: ]{ekolemen@princeton.edu}
\email[]{ekolemen@pppl.gov}
\affiliation{Department of Mechanical and Aerospace Engineering, Princeton University, Princeton, New Jersey, 08540, USA}
\affiliation{Princeton Plasma Physics Laboratory, Princeton, New Jersey, 08540, USA}

\date{\today}

\begin{abstract}
Free-surface liquid metal flows are a leading candidate for the plasma-facing components of future fusion reactors, but existing transient, three-dimensional, free-surface MHD solvers rely on the inductionless approximation, in which the induced magnetic field is neglected. This paper extends the open-source solver \texttt{FreeMHD} [B.~Wynne \textit{et al.}, Phys.~Plasmas \textbf{32}, 013907 (2025)] beyond that approximation, resolving the induced field self-consistently with a vector-potential formulation that enforces $\nabla\cdot\boldsymbol{B}=0$ by construction while preserving the original multi-region, two-phase framework. It is verified against the analytical Shercliff and Hunt duct flows, against flows driven by a time-varying applied field, and against the deformation of a free liquid metal jet crossing a non-uniform field, and validated against free-surface height measurements from the LMX-U experiment. The experiment validates the overall free-surface solution rather than finite-$R_m$ effects, which the transient-field cases verify. To our knowledge this is the first open-source, fully three-dimensional free-surface liquid metal solver to resolve the evolution of the induced magnetic field, providing a basis for modeling the finite magnetic Reynolds number conditions expected in large-scale, transient fusion events.
\end{abstract}

\maketitle

\section{Introduction}\label{introduction}

Electrically conductive flows are relevant to a wide range of industrial applications such as electromagnetic braking in metallurgy,\cite{HaiqiJMPT08} liquid metal batteries for grid-scale storage,\cite{ZhangEES21} electromagnetic pumping\cite{XiangTSEP24} and propulsion systems,\cite{CortazarAMT25} and magnetically controlled melt flows in semiconductor crystal growth.\cite{YangIJHFF07} Among these, free-surface liquid metal flows have attracted particular attention as a leading candidate for plasma-facing components (PFCs) in future fusion reactors. A wide variety of free-surface liquid metal divertor concepts have been proposed,\cite{MiyazawaFED17,AlegreJFE20,FisherNF20,RindtFED21,SuarezFED21,SaenzNF22} motivated by the clear advantages that liquid metals offer over solid wall materials.
 
Chief among these advantages is the ability of a flowing liquid metal to effectively advect the intense thermal energy deposited on the divertor surface, provided the flow speed is sufficient. Since the liquid metal is continuously replenished, there is no concern for the irreversible structural damage that limits the lifetime of solid PFCs. However, the primary obstacle that must be overcome is the $\boldsymbol{j}\times\boldsymbol{B}$ drag force, which retards the bulk flow of any electrically conducting fluid moving through a strong magnetic field. In typical tokamak reactor designs, divertors are expected to operate under magnetic fields of order $B_0\sim10\,\mathrm{T}$, with peak heat fluxes of $Q\sim10\,\mathrm{MW\cdot m^{-2}}$ and neutron wall loadings of $\Gamma_n\sim1\,\mathrm{MW\cdot m^{-2}}$.\cite{SergeevNF15,PintsukFED22,ZhangAA25} These extreme conditions are difficult to reproduce even in the most advanced experimental facilities currently in operation, making numerical modeling an indispensable tool in the design of liquid metal divertors.
 
Extensive efforts have been made to develop magnetohydrodynamics (MHD) codes for electrically conductive flows. However, most existing codes are limited in important ways: they assume steady-state conditions, are restricted to two-dimensional dynamics, or can only treat internal flows with no free surface. Only a handful of codes are capable of modeling transient, three-dimensional, free-surface flows,\cite{MorleyFED04,SamulyakJCP07,SirianoFED24} and without exception they all rely on the inductionless approximation, i.e., the low magnetic Reynolds number approximation ($R_m\equiv\mu_0\sigma UL \ll 1$). Here, $\mu_0$ is the magnetic permeability of free space, $\sigma$ is the electric conductivity, and $U$ and $L$ are the characteristic speed and length, respectively.
 
Under this approximation, the induced magnetic field generated by the liquid metal is assumed negligible compared to the externally applied field, so the induction equation reduces to a diffusion equation and the magnetic field is effectively frozen at its externally prescribed value. This assumption has been invoked in most fusion-relevant studies on the grounds that steady-state reactor operation yields a small $R_m$.\cite{AlSalamiKU22,SuarezITPS22} However, a conclusive demonstration of its validity has yet to be presented.

A complicating subtlety, specific to the free-surface configurations of interest here, is the choice of characteristic length $L$. For closed-channel liquid metal MHD flows the convention is well established, $L$ being taken as the half-width of the duct in the magnetic field direction, and no ambiguity arises. Free-surface flows are less settled, because the geometry itself is part of the solution: the film depth is not fixed a priori, and open-channel practice commonly characterizes the flow through the hydraulic diameter, and hence the wetted perimeter. In a thin flowing film the depth, the spanwise width, and the streamwise extent over which the flow interacts with the field can differ by one to two orders of magnitude, so estimates of $R_m$ built on these different scales differ accordingly. 

More critically, the inductionless approximation is least defensible precisely in the scenarios of greatest engineering concern: large-scale electromagnetic events such as giant edge-localized modes (ELMs), vertical displacement events (VDEs), and major or minor plasma disruptions can drive strong transient coupling between the plasma and the liquid metal, conditions under which the low-$R_m$ assumption would be seriously compromised.\cite{SmolentsevFED10,KawczynskiFED16,KawczynskiPoF18,SmolyanovPRE25} Furthermore, alternative fusion reactor concepts involving large-scale liquid metal flows are increasingly prominent, and these flows are expected to reach the finite-$R_m$ regime from the outset.\cite{LevittPoP23}
 
There is therefore a clear and pressing need for free-surface MHD solvers that resolve the evolution of the induced magnetic field without invoking the inductionless approximation. The broader MHD community has begun moving in this direction: Endeve~\textit{et al.}\cite{EndeveFED26} recently developed a full-induction solver for internal liquid metal flows in blanket duct geometries. Within the \texttt{OpenFOAM} ecosystem specifically, several related capabilities have been reported. Blishchik~\textit{et al.}\cite{BlishchikIJHFF21} presented an extensive numerical benchmark spanning conjugate and multiphase volume-of-fluid MHD configurations under the inductionless approximation, and Su\'arez and Smolentsev\cite{SuarezFED25} recently developed three simulation tools of various dimensionality, a three-dimensional volume-of-fluid model and reduced two- and one-dimensional counterparts, for open-surface liquid metal flows in plasma-facing components, likewise inductionless. Going beyond the inductionless approximation, Suponitsky~\textit{et al.}\cite{SuponitskyFluids22} developed a compressible two-phase free-surface MHD solver that evolves the magnetic field self-consistently in terms of its toroidal component and poloidal flux function, and subsequently validated it against measured liner trajectories and magnetic-field amplification in two lithium-liner implosion experiments;\cite{SuponitskyFluids25} that solver is, however, restricted to axisymmetric problems and is not openly released. To our knowledge, no open-source solver has yet combined fully three-dimensional free-surface capability with a self-consistently resolved induced magnetic field in a single validated framework.
 
To bridge this gap, we extend the open-source, multi-domain, two-phase solver \texttt{FreeMHD}\cite{AlSalamiKU22,WynnePoP25} beyond the inductionless approximation by resolving the evolution of the induced magnetic field directly. The induction equation is solved using a vector potential ($\boldsymbol{A}$) formulation, which naturally preserves the solenoidal constraint $\nabla\cdot\boldsymbol{B}=0$ by construction, avoiding the divergence errors that have plagued many MHD codes. The resulting solver retains the full multi-region, two-phase architecture of \texttt{FreeMHD} while removing the structural assumption that the induced field is negligible, which is a prerequisite for modeling the finite-$R_m$ regime. To the best of our knowledge, it is the first open-source solver to resolve the evolution of the induced magnetic field in fully three-dimensional free-surface liquid metal flows without invoking the inductionless approximation; its novelty thus lies in the public release and the combination of these capabilities within a single experimentally validated framework. Beyond verification and validation, the resolved solution already reveals, even at the low $R_m$ of the validation experiment, induced-field and surface-current structures in a realistic free-surface configuration that are inaccessible to inductionless solvers (Sec.~\ref{sec:experimental_validation}).
 
This paper is organized as follows. The governing equations are presented in Sec.~\ref{governing equations}. The numerical implementation is detailed in Sec.~\ref{numerical implementation}. Verification and validation results are reported in Sec.~\ref{verification and validation}, and conclusions are drawn in Sec.~\ref{conclusion}.

\section{Governing Equations}\label{governing equations}

\subsection{Resistive MHD equations}

The incompressible resistive MHD governing equations solved by the code are as follows. First, the incompressibility condition states
\begin{equation}\label{incompressibility}
    \nabla\cdot \boldsymbol{u}=0.
\end{equation}
Next, the Navier--Stokes equation for momentum reads
\begin{equation}\label{Navier Stokes}
    \frac{\partial}{\partial t}\left(\rho \boldsymbol{u}\right)+\nabla\cdot(\rho \boldsymbol{u}\otimes \boldsymbol{u})=-\nabla p+\nabla\cdot\boldsymbol{\tau} +\rho \boldsymbol{g}+\boldsymbol{j}\times \left(\nabla\times \boldsymbol{A}\right)+\boldsymbol{F}_{\mathrm{st}}.
\end{equation}
The induction equation is given by
\begin{equation}\label{induction}
    \frac{\partial}{\partial t}\boldsymbol{A}=-\nabla\phi+\boldsymbol{u}\times\left(\nabla\times \boldsymbol{A}\right)+\frac{1}{\mu_0\sigma}\nabla^2 \boldsymbol{A},
\end{equation}
where the Coulomb gauge is imposed.
The scalar potential $\phi$ is solved with the Poisson equation
\begin{equation}\label{Poisson}
    \nabla\cdot\left(\sigma\nabla\phi\right)=\nabla\cdot\left[\sigma \boldsymbol{u}\times\left(\nabla\times \boldsymbol{A}\right)-\sigma\frac{\partial\boldsymbol{A}}{\partial t}\right],
\end{equation}
where the time-varying term can be ignored if the conductivity is spatially uniform. The current density obeys Ampere's law
\begin{equation}\label{Amperes law}
    \boldsymbol{j}=\frac{1}{\mu_0}\nabla\times\left(\nabla\times \boldsymbol{A}\right).
\end{equation}
Here, $\boldsymbol{u}$ is the flow velocity, $\rho$ is the mass density, $p$ is the pressure, $\boldsymbol{\tau}$ is the stress tensor, $\boldsymbol{g}$ is the gravitational acceleration, $\boldsymbol{j}$ is the current density, $\boldsymbol{A}$ is the vector potential, and $\boldsymbol{F}_{\mathrm{st}}$ is the surface tension force. Surface tension modeling is described in a later section. The heat transport equation has recently been implemented in \texttt{FreeMHD}, but will not be used in this study and will be reported in a separate publication after sufficient verification and validation.

Although the expression for current density shown in Eq.~(\ref{Amperes law}) is valid, the second-order spatial derivative in this expression makes it vulnerable to numerical artifacts and noise. Hence, in practice, the following form based on Ohm's law is preferred when evaluating the current density:
\begin{equation}\label{Ohms law}
    \boldsymbol{j}=\sigma(\boldsymbol{E}+\boldsymbol{u}\times\boldsymbol{B})=\sigma\left[-\nabla\phi-\frac{\partial \boldsymbol{A}}{\partial t}+\boldsymbol{u}\times(\nabla\times\boldsymbol{A})\right].
\end{equation}

\subsection{Induced-field formulation}

The main extension reported in this paper revolves around Eq.~(\ref{induction}). Previously, a quasi-steady-state approximation, also known as the inductionless approximation, was used to set $\partial\boldsymbol{A}/\partial t=0$, thus allowing $\boldsymbol{B}\approx \boldsymbol{B}_{\mathrm{ext}}$ where $\boldsymbol{B}_{\mathrm{ext}}$ is the externally applied magnetic field. This approximation was justified in the low-$R_m$ regime, which can be defined as the ratio of the magnetic advection term [second term on the right-hand side of Eq.~(\ref{induction})] to the magnetic diffusion term [last term of Eq.~(\ref{induction})] (the first term is a gauge term). For $R_m\ll1$, the induction equation becomes a diffusion equation, resulting in the magnetic field quickly relaxing toward the value determined solely by the boundary conditions.

For the target applications considered here, the magnetic field is split into two components, i.e., $\boldsymbol{B}=\nabla\times \boldsymbol{A}_{\mathrm{ind}}+\boldsymbol{B}_{\mathrm{ext}}$, thus replacing $\nabla\times\boldsymbol{A}$ in Eqs.~(\ref{Navier Stokes})--(\ref{Ohms law}) with $\nabla\times\boldsymbol{A}_{\mathrm{ind}}+\boldsymbol{B}_{\mathrm{ext}}$. This split facilitates numerical modeling of fusion reactor applications, where a strong external magnetic field is applied either by coils or by plasma, and the field induced by the liquid metal can be evolved self-consistently. This external magnetic field is curl-free in the computational region, i.e., $\nabla\times \boldsymbol{B}_{\mathrm{ext}}=0$. The external field is treated as magnetic field $\boldsymbol{B}$ for ease of preparing code input, whereas the induced field is solved in the vector potential formulation to circumvent the numerical issues arising from enforcing the divergence-free constraint on the magnetic field $\nabla\cdot \boldsymbol{B}=0$. This is possible since divergence of a curl of any vector is, by definition, identically zero ($\nabla\cdot\left(\nabla\times \boldsymbol{A}\right)\equiv0$).

\section{Numerical Implementation}\label{numerical implementation}

This section describes how the governing equations of Sec.~\ref{governing equations} are discretized and solved within the present framework. The finite-volume discretization, the volume-of-fluid free-surface treatment, and the pressure--velocity coupling are inherited from \texttt{FreeMHD}; the electromagnetic solution, comprising the coupled induced vector potential and scalar potential system solved within and across the multi-region domains (Secs.~\ref{fluid mhd}--\ref{boundary conditions}), constitutes the principal extension reported here.

\subsection{Numerical framework}\label{numerical framework}

The solver is built upon the finite-volume infrastructure of OpenFOAM~\cite{WellerCP98} (v2206) and is implemented as a custom multi-region solver derived from its two-phase volume-of-fluid solver families. The governing equations of Sec.~\ref{governing equations} are discretized with the finite-volume method on a collocated variable arrangement, in which all primary fields (velocity $\boldsymbol{u}$, pressure $p$, vector potential $\boldsymbol{A}$, scalar potential $\phi$, and phase fraction $\alpha$, with the last of these defined in Sec.~\ref{free surface modeling}) are stored at cell centroids and integrated over control volumes, with the corresponding fluxes evaluated on cell faces. This arrangement enforces discrete conservation and is compatible with arbitrary unstructured polyhedral meshes.

Spatial discretization is second-order: diffusive terms employ central differencing, while advective terms use bounded interpolation schemes to preserve numerical stability and the boundedness of transported quantities such as the phase fraction. The equations are advanced in time with a segregated, sequential procedure, and the time step is adapted at runtime to satisfy prescribed Courant-number limits on both the bulk flow and the advecting interface, subject to a fixed maximum-step ceiling.

One of the defining features of the solver is its multi-region architecture. The computational domain is partitioned into an arbitrary number of subdomains of three types: electrically conducting fluid regions, in which the full set of two-phase MHD equations is solved; solid regions, representing structural components such as walls, which can have arbitrary finite electric conductivity; and vacuum regions, which carry only the electromagnetic fields. Each region is meshed independently and has its own set of fields, and the regions communicate exclusively across their shared interfaces, which are coupled through mapped boundary patches (see Sec.~\ref{boundary conditions}). Two-phase capability is realized within the fluid region: a fluid region contains \emph{both} phases, the liquid metal and the gas, on a single shared mesh, with the free surface captured by the volume-of-fluid method of Sec.~\ref{free surface modeling} rather than by a region interface (the region decomposition is illustrated in Fig.~\ref{fig:regions}).

Within the fluid regions, pressure--velocity coupling is handled by the PIMPLE algorithm, a hybrid of the PISO~\cite{IssaJCP86} and SIMPLE~\cite{CarettoProc73} methods. The electromagnetic equations are solved on the same mesh as the flow variables, so that the two-way coupling between flow and field (the Lorentz force in the momentum equation and the motional electromotive force in the induction equation) is realized directly through source terms rather than through any separate field-transfer step. Each time step comprises a fixed number of outer (PIMPLE) correctors. Within every corrector the regions are visited in sequence (fluid, then solid, then vacuum), with the solid regions optionally sub-iterated to tighten the interfacial coupling. The full sequence is detailed in Sec.~\ref{solution algorithm} and summarized in Algorithm~\ref{alg:timestep}.

\subsection{Free-surface capturing and phase-dependent properties}
\label{free surface modeling}

Free-surface dynamics are captured with the algebraic volume-of-fluid (VOF) method,\cite{DeshpandeCSD12} which represents the two immiscible phases (liquid metal and the surrounding gas) through a single indicator function $I(\boldsymbol{x},t)$ that equals unity in the liquid metal and zero in the gas. Its control-volume average defines the phase fraction \begin{equation}\label{alpha definition}
\alpha = \frac{1}{\Omega}\int_\Omega I(\boldsymbol{x},t)\,dV,
\end{equation}
where $\Omega$ is the cell volume, so that $\alpha$ varies smoothly between zero and one across the resolved interface. The phase fraction is transported by
\begin{equation}\label{alpha continuity}
\frac{\partial \alpha}{\partial t} + \nabla\cdot\left(\alpha\boldsymbol{u}\right)
= 0,
\end{equation}
which is solved with the MULES limiter together with an interface-compression flux that counteracts numerical smearing while keeping $\alpha$ bounded within $[0,1]$. To respect the interface Courant limit independently of the bulk-flow time step, Eq.~(\ref{alpha continuity}) is advanced over sub-cycles within each time step.

Surface tension is incorporated through the continuum surface force (CSF) model,\cite{BrackbillJCP92} which recasts the interfacial force as the volumetric source $\boldsymbol{F}_{\mathrm{st}}$ that appears in the momentum equation [Eq.~(\ref{Navier Stokes})],
\begin{equation}\label{surface tension}
\boldsymbol{F}_{\mathrm{st}} = \sigma_{\mathrm{st}}\,\kappa\,\nabla\alpha,
\end{equation}
with surface-tension coefficient $\sigma_{\mathrm{st}}$ and interface curvature
\begin{equation}\label{curvature}
\kappa = \nabla\cdot\left(\frac{\nabla\alpha}{|\nabla\alpha|}\right).
\end{equation}

Material properties are assigned cell by cell from the phase fraction. To guard against the small over- and undershoots that algebraic VOF can produce, the fraction is first clamped,
\begin{equation}\label{alpha clamp}
\alpha_c = \min\!\left(\max(\alpha,0),\,1\right),
\end{equation}
and each phase-dependent property is then evaluated as a linear blend of its single-phase values. The property of central importance to the electromagnetic problem is the electric conductivity,
\begin{equation}\label{alpha interpolation}
\sigma = \alpha_c\,\sigma_1 + (1-\alpha_c)\,\sigma_2,
\end{equation}
where subscripts $1$ and $2$ denote the liquid metal and gas phases. Because the conductivity contrast between the phases spans several orders of magnitude, the linear blend of Eq.~(\ref{alpha interpolation}) assigns the cells of the resolved interface an intermediate conductivity, so that the transition layer behaves as a thin, partially conducting sheet rather than as a discontinuity. The consequence is that the electromagnetic solution inherits the finite thickness of the volume-of-fluid interface: currents that would physically close in a surface layer of vanishing thickness are instead distributed over the two to three cells that the interface occupies. This is an intrinsic feature of algebraic interface capturing rather than of the electromagnetic formulation, and it is benign as long as the interface is thin compared with the layer of fluid carrying the current; the extent to which that condition is met in the present calculations is quantified in Sec.~\ref{sec:experimental_validation}. The magnetic permeability is assumed to be equal to the vacuum permeability $\mu_0$ in every phase and region treated in this work and is therefore held constant throughout the domain. The conductivity, by contrast, falls by many orders of magnitude across the interface, from the large value of the liquid metal to the negligible value of the gas. This conductivity contrast ties the electromagnetic solution directly to the instantaneous interface position, since the current density evaluated in Sec.~\ref{fluid mhd} and the induced field both depend on where the conducting phase terminates. The free surface and the electromagnetic fields therefore evolve as a two-way coupled system.

This two-way coupling places a demand on the mesh that is peculiar to free-surface MHD. The conductivity gradient that supports the interfacial current lies wherever the interface happens to be, and since that position is not known before the computation, the mesh cannot be graded toward it in the way it can be graded toward a wall. The present solver does not adapt the mesh to follow the interface as it moves. Instead, the band of the domain in which the interface is expected to lie is resolved uniformly, its width taken from the expected operating envelope of the flow and the computed interface confirmed a posteriori to remain within it. The sensitivity of the results to that resolution is reported in Sec.~\ref{verification and validation}.

\subsection{Magnetic field solution in the fluid region}\label{fluid mhd}

Within each fluid region the electromagnetic state is described by the scalar potential $\phi$ and the induced magnetic vector potential $\boldsymbol{A}_{\mathrm{ind}}$, from which the magnetic field is reconstructed as
\begin{equation}\label{B reconstruction}
\boldsymbol{B} = \nabla\times\boldsymbol{A}_{\mathrm{ind}} + \boldsymbol{B}_{\mathrm{ext}},
\end{equation}
where $\boldsymbol{B}_{\mathrm{ext}}$ is the prescribed external field and $\nabla\times\boldsymbol{A}_{\mathrm{ind}}$ is the induced field. The formulation requires only that $\boldsymbol{B}_{\mathrm{ext}}$ be solenoidal, the current density in the conductor being obtained from Ohm's law rather than from the curl of the total field; it need not itself be a vacuum field.

When the applied field varies in time it drives current directly, through the electromotive force associated with its own vector potential $\boldsymbol{A}_{\mathrm{ext}}$, defined by $\boldsymbol{B}_{\mathrm{ext}}=\nabla\times\boldsymbol{A}_{\mathrm{ext}}$. The total vector potential is $\boldsymbol{A}=\boldsymbol{A}_{\mathrm{ind}}+\boldsymbol{A}_{\mathrm{ext}}$, and it is this total that enters Ohm's law [Eq.~(\ref{Ohms law})] and charge conservation [Eq.~(\ref{Poisson})]. Only $\boldsymbol{A}_{\mathrm{ind}}$ is solved for. The rate $\partial\boldsymbol{A}_{\mathrm{ext}}/\partial t$ is prescribed analytically alongside $\boldsymbol{B}_{\mathrm{ext}}$, so that $\nabla\times\boldsymbol{A}_{\mathrm{ext}}=\boldsymbol{B}_{\mathrm{ext}}$ holds by construction rather than through a further solve. The gauge in which it is prescribed is problem-specific; the two used here are given in Sec.~\ref{sec:transient_verification}. Where the applied field is static this term vanishes identically and each expression below reduces to the static-field form used in the remaining cases.

The two potentials are governed by the induction and charge-conservation equations [Eqs.~(\ref{induction}) and~(\ref{Poisson})] of Sec.~\ref{governing equations}. In the finite-volume implementation the induced vector potential is advanced in the conductivity-weighted form
\begin{equation}\label{A solved}
\left(\sigma\,\frac{\partial}{\partial t} - \frac{1}{\mu_0}\nabla^2\right)
\boldsymbol{A}_{\mathrm{ind}}
=- \sigma\nabla\phi + \sigma\,\bigl[\boldsymbol{u}\times\left(\nabla\times\boldsymbol{A}_{\mathrm{ind}}+\boldsymbol{B}_{\mathrm{ext}}\right)\bigr]
{}- \sigma\,\frac{\partial\boldsymbol{A}_{\mathrm{ext}}}{\partial t},
\end{equation}
obtained by combining Ohm's law with Ampere's law under the Coulomb gauge, while the scalar potential is obtained from the Poisson equation [Eq.~(\ref{Poisson})], whose solution is pinned at a reference cell to remove the additive-constant indeterminacy.

Eqs.~(\ref{A solved}) and~(\ref{Poisson}) are mutually coupled: the scalar potential source depends on $\boldsymbol{A}_{\mathrm{ind}}$ through $\partial\boldsymbol{A}_{\mathrm{ind}}/\partial t$ and on $\nabla\times\boldsymbol{A}_{\mathrm{ind}}$, while the vector potential source depends on $\phi$. They are resolved by a Picard iteration that, within each call, solves the Poisson equation [Eq.~(\ref{Poisson})] for $\phi$, then Eq.~(\ref{A solved}) for $\boldsymbol{A}_{\mathrm{ind}}$, updates $\boldsymbol{B}$ through Eq.~(\ref{B reconstruction}), and repeats until the larger of the two initial residuals falls below a fixed tolerance or a prescribed iteration count is reached. The time-derivative term $\partial\boldsymbol{A}_{\mathrm{ind}}/\partial t$ that appears in the scalar potential source is evaluated once, from the field at the time-step level, and held fixed throughout the iteration. This is essential for stability: with the small time steps required by the interface Courant limit, re-evaluating $\partial\boldsymbol{A}_{\mathrm{ind}}/\partial t$ at every iterate would amplify the inter-iterate change in $\boldsymbol{A}_{\mathrm{ind}}$ by a factor $1/\Delta t$ in the Poisson source, driving the iteration into a sustained limit cycle rather than convergence. Once the loop has converged, $\partial\boldsymbol{A}_{\mathrm{ind}}/\partial t$ is re-evaluated from the updated field for use in the current-density evaluation. No such treatment is needed for $\partial\boldsymbol{A}_{\mathrm{ext}}/\partial t$, which is known in closed form at the current time and is evaluated once per time step, exactly rather than as a lagged difference.

The electromagnetic source terms are assembled on cell faces and only then transferred to cell centers, rather than being evaluated pointwise at cell centers. Let $[\,\cdot\,]_f$ denote face interpolation, $\boldsymbol{S}_f$ the face-area vector, $\sigma_f$ the face-interpolated conductivity, and $\nabla_{\!f}\phi$ the face-normal gradient of $\phi$. The source of the scalar potential equation [Eq.~(\ref{Poisson})] is built as the face flux
\begin{equation}\label{psiub}
\Phi_\sigma = \left[\sigma\left(\boldsymbol{u}\times\left(\nabla\times\boldsymbol{A}_{\mathrm{ind}}+\boldsymbol{B}_{\mathrm{ext}}\right)
- \frac{\partial \boldsymbol{A}_{\mathrm{ind}}}{\partial t}
{}- \frac{\partial \boldsymbol{A}_{\mathrm{ext}}}{\partial t}\right)\right]_f \cdot
\boldsymbol{S}_f,
\end{equation}
whose divergence forms the right-hand side of the Poisson equation, while the source of the vector potential equation [right-hand side of Eq.~(\ref{A solved})] is assembled from the same face quantities and reconstructed to cell centers,
\begin{equation}\label{sAf}
\boldsymbol{s}_{\boldsymbol{A}} = \mathrm{reconstruct}\!\left(
- \sigma_f\,(\nabla_{\!f}\phi)\,|\boldsymbol{S}_f|
+ \left[\sigma\left(\boldsymbol{u}\times(\nabla\times\boldsymbol{A}_{\mathrm{ind}}+\boldsymbol{B}_{\mathrm{ext}})
{}- \frac{\partial\boldsymbol{A}_{\mathrm{ext}}}{\partial t}\right)\right]_f\cdot\boldsymbol{S}_f
\right),
\end{equation}
where $\mathrm{reconstruct}(\cdot)$ denotes the cell-center vector recovered from a set of face fluxes. The induced time-derivative term $\partial\boldsymbol{A}_{\mathrm{ind}}/\partial t$ is absent from Eq.~(\ref{sAf}) because it is treated implicitly through the transient term on the left-hand side of Eq.~(\ref{A solved}), whereas it appears explicitly in Eq.~(\ref{psiub}). Although the Coulomb gauge sets $\nabla\cdot\boldsymbol{A}_{\mathrm{ind}}=0$, this contribution to the charge-conservation source does not vanish, because the conductivity is non-uniform across the phases: $\nabla\cdot(\sigma\,\partial\boldsymbol{A}_{\mathrm{ind}}/\partial t)=\nabla\sigma\cdot\partial\boldsymbol{A}_{\mathrm{ind}}/\partial t$, which is supported on the interface where $\nabla\sigma\neq 0$. The term is therefore retained in the fluid regions; in the solid regions it vanishes identically and is omitted (Sec.~\ref{non fluid mhd}). The external rate $\partial\boldsymbol{A}_{\mathrm{ext}}/\partial t$ cannot be absorbed implicitly in the same way, the transient operator of Eq.~(\ref{A solved}) acting on $\boldsymbol{A}_{\mathrm{ind}}$ alone, and is therefore carried explicitly in Eq.~(\ref{psiub}), Eq.~(\ref{sAf}) and Eq.~(\ref{jn}) alike, so that the same electromotive force appears in the charge-conservation source, in the vector-potential source and in the current density. Finally, because the $-\sigma\nabla\phi$ contribution to Eq.~(\ref{sAf}) is built from the same face-normal gradient that the Poisson equation solves for, and the motional contribution from the same face interpolation as in Eq.~(\ref{psiub}), the current implied by the vector potential source is consistent, face by face, with the current entering charge conservation, which suppresses the odd-even decoupling between $\phi$ and $\boldsymbol{A}_{\mathrm{ind}}$ that would otherwise generate spurious, non-solenoidal currents.

Consistent with this construction, the current density is evaluated from the Ohm's law face flux
\begin{equation}\label{jn}
j_n = -\,\sigma_f\,(\nabla_{\!f}\phi)\,|\boldsymbol{S}_f|
+ \left[\sigma\left(\boldsymbol{u}\times(\nabla\times\boldsymbol{A}_{\mathrm{ind}}+\boldsymbol{B}_{\mathrm{ext}})
- \frac{\partial\boldsymbol{A}_{\mathrm{ind}}}{\partial t}
{}- \frac{\partial\boldsymbol{A}_{\mathrm{ext}}}{\partial t}\right)\right]_f\cdot
\boldsymbol{S}_f,
\end{equation}
which is reconstructed to the cell-centered current $\boldsymbol{j}=\mathrm{reconstruct}(j_n)$. The Lorentz force entering the momentum equation [Eq.~(\ref{Navier Stokes})] is assembled in a conservative form built from the current flux $j_n$ and the face-interpolated field, thereby conserving momentum discretely.

This construction is the vector-potential counterpart of the consistent and conservative schemes of Ni~\textit{et al.}\cite{NiJCP07a,NiJCP07b} for the inductionless formulation. Those schemes evaluate the current density as a face flux from the same discrete gradient operator that the electric-potential equation inverts, which conserves charge to machine accuracy and avoids checkerboard decoupling between $\phi$ and $\boldsymbol{j}$, and they assemble the Lorentz force from those fluxes rather than from cell-centered products, which conserves momentum discretely. Equations~(\ref{psiub})--(\ref{jn}) implement both: $\Phi_\sigma$, $\boldsymbol{s}_{\boldsymbol{A}}$ and $j_n$ are built from a single face-flux expression, and the force is assembled from $j_n$ and the face-interpolated field. Two differences follow from resolving the induced field. The electromotive contribution to each face flux is $\boldsymbol{u}\times(\nabla\times\boldsymbol{A}_{\mathrm{ind}}+\boldsymbol{B}_{\mathrm{ext}})-\partial\boldsymbol{A}_{\mathrm{ind}}/\partial t-\partial\boldsymbol{A}_{\mathrm{ext}}/\partial t$ rather than $\boldsymbol{u}\times\boldsymbol{B}_{\mathrm{ext}}$ alone, so consistency extends to the vector-potential equation, whose source must be built from the same faces as $\Phi_\sigma$; this is the reason for the reconstruction in Eq.~(\ref{sAf}). And the applied field need not be constant, in space or in time, for the assembly to conserve momentum, since the field entering it is the reconstructed total field of Eq.~(\ref{B reconstruction}), whereas Ref.~\onlinecite{NiJCP07b} splits the Lorentz force into globally and locally conservative parts.

\subsection{Magnetic field solution in non-fluid regions}\label{non fluid mhd}

The non-fluid regions carry no flow, so the motional term $\sigma(\boldsymbol{u}\times\boldsymbol{B})$ is absent throughout. They are divided into two types, conducting and insulating regions, which are treated differently. Each conducting solid region carries its own, arbitrary finite electrical conductivity. Walls of spatially non-uniform conductivity are handled by decomposing the wall into multiple solid regions, each with its own conductivity, coupled through the interface conditions of Sec.~\ref{boundary conditions}.

In the conducting solid regions the scalar and induced vector potentials are advanced by the same coupled iterative procedure used in the fluid (Sec.~\ref{fluid mhd}), including the frozen-$\partial\boldsymbol{A}_{\mathrm{ind}}/\partial t$ treatment and the face-flux assembly of the source terms, with two simplifications. First, because the conductivity is uniform within a solid, the time-derivative contribution to the charge-conservation source vanishes identically, as anticipated in Sec.~\ref{fluid mhd}, and the scalar potential obeys the source-free equation
\begin{equation}\label{solid phi}
\nabla\cdot(\sigma\nabla\phi) = 0.
\end{equation}
Second, with the motional term removed, the source of the vector potential equation [Eq.~(\ref{A solved})] reduces to $-\sigma\nabla\phi$, and the current density follows from Eq.~(\ref{jn}) with the same omission,
\begin{equation}\label{solid jn}
j_n = -\,\sigma_f\,(\nabla_{\!f}\phi)\,|\boldsymbol{S}_f|
- \left[\sigma\,\frac{\partial\boldsymbol{A}_{\mathrm{ind}}}{\partial t}\right]_f\cdot
\boldsymbol{S}_f.
\end{equation}
The induced vector potential retains its transient and diffusive terms, so a solid region behaves as a conductor in which eddy currents can be induced rather than as a passive Laplace medium.

The insulating solid and vacuum regions contain no current. The induced vector potential there satisfies the vector Laplace equation
\begin{equation}\label{vacuum A}
\nabla^2\boldsymbol{A}_{\mathrm{ind}} = 0,
\end{equation}
which is solved iteratively to a fixed residual tolerance, and the magnetic field follows as $\boldsymbol{B} = \nabla\times\boldsymbol{A}_{\mathrm{ind}}+\boldsymbol{B}_{\mathrm{ext}}$. A vacuum region is included so that the induced field can decay smoothly toward a prescribed far-field condition on its outer boundary; the closure used for that boundary is described in Sec.~\ref{boundary conditions}.

\subsection{Boundary and interface conditions}\label{boundary conditions}

Because each region is meshed independently, the regions exchange information only across their shared interfaces. Every interface condition is implemented as a mapped boundary patch that retrieves the corresponding field values from the adjacent region and imposes the appropriate continuity or flux condition. The electromagnetic conditions are summarized in Table~\ref{bc table} and described below.

The induced vector potential $\boldsymbol{A}_{\mathrm{ind}}$ is required to be continuous across every interior interface, which is enforced through a mixed condition that sets the interface value to the distance-weighted average of the adjacent cell values. Its time derivative $\partial\boldsymbol{A}_{\mathrm{ind}}/\partial t$ is treated consistently: rather than being coupled independently, it is set on each interface to the discrete time derivative of the already-continuous $\boldsymbol{A}_{\mathrm{ind}}$ face value, so that it inherits the continuity of $\boldsymbol{A}_{\mathrm{ind}}$ without a separate neighbor exchange. The elliptic problem for $\boldsymbol{A}_{\mathrm{ind}}$ is closed by enclosing the domain in vacuum and imposing $\boldsymbol{A}_{\mathrm{ind}}=\boldsymbol{0}$ on its outer surface, which provides the far-field anchor required for a unique solution; the more general exterior treatment is deferred to a separate publication.

The scalar potential $\phi$ is solved only in the conducting regions, and its interface conditions follow from the continuity of the normal current density. With $\hat{\boldsymbol{n}}$ denoting the interface normal, the normal current is
\begin{equation}\label{normal current}
j_n = \sigma\!\left(-\frac{\partial\phi}{\partial n}
+ \left(\boldsymbol{u}\times(\nabla\times\boldsymbol{A}_{\mathrm{ind}}+\boldsymbol{B}_{\mathrm{ext}})
- \frac{\partial\boldsymbol{A}_{\mathrm{ind}}}{\partial t}
{}- \frac{\partial\boldsymbol{A}_{\mathrm{ext}}}{\partial t}\right)\cdot\hat{\boldsymbol{n}}
\right).
\end{equation}
Then, two cases arise. At conductor--conductor interfaces $j_n$ is required to be continuous, imposed as a mixed condition weighted by the face conductance $\sigma/\delta$, with $\delta$ the cell-to-face distance, so that the current passing from one conductor into the other is conserved. At conductor--insulator interfaces the normal current must vanish ($j_n=0$) since no current can enter the non-conducting region; this is realized by fixing the normal gradient of $\phi$ to the local electromotive projection, $\partial\phi/\partial n = [\boldsymbol{u}\times(\nabla\times\boldsymbol{A}_{\mathrm{ind}}+\boldsymbol{B}_{\mathrm{ext}})- \partial\boldsymbol{A}_{\mathrm{ind}}/\partial t{}- \partial\boldsymbol{A}_{\mathrm{ext}}/\partial t]\cdot\hat{\boldsymbol{n}}$, with $\boldsymbol{u}=\boldsymbol{0}$ in the solid regions.

\begin{table}
\caption{\label{bc table}Electromagnetic boundary and interface conditions for the induced vector potential, its time derivative, and the scalar potential. Entries marked n/a are not solved in the adjacent region. (Sec.~\ref{boundary conditions})}
\begin{ruledtabular}
\begin{tabular}{lccc}
Interface & $\boldsymbol{A}_{\mathrm{ind}}$ & $\partial\boldsymbol{A}_{\mathrm{ind}}/\partial t$
& $\phi$ \\
\hline
Conductor--conductor & continuous & continuous & $j_n$ continuous \\
Conductor--insulator    & continuous & continuous & $j_n=0$ \\
Insulator--insulator       & continuous & continuous & n/a \\
Exterior             & $\boldsymbol{A}_{\mathrm{ind}}=\boldsymbol{0}$ & $\boldsymbol{0}$
& n/a \\
\end{tabular}
\end{ruledtabular}
\end{table}

\subsection{Overall solution algorithm}\label{solution algorithm}

\begin{figure}[htbp]
  \centering
  \includegraphics[width=0.6\linewidth]{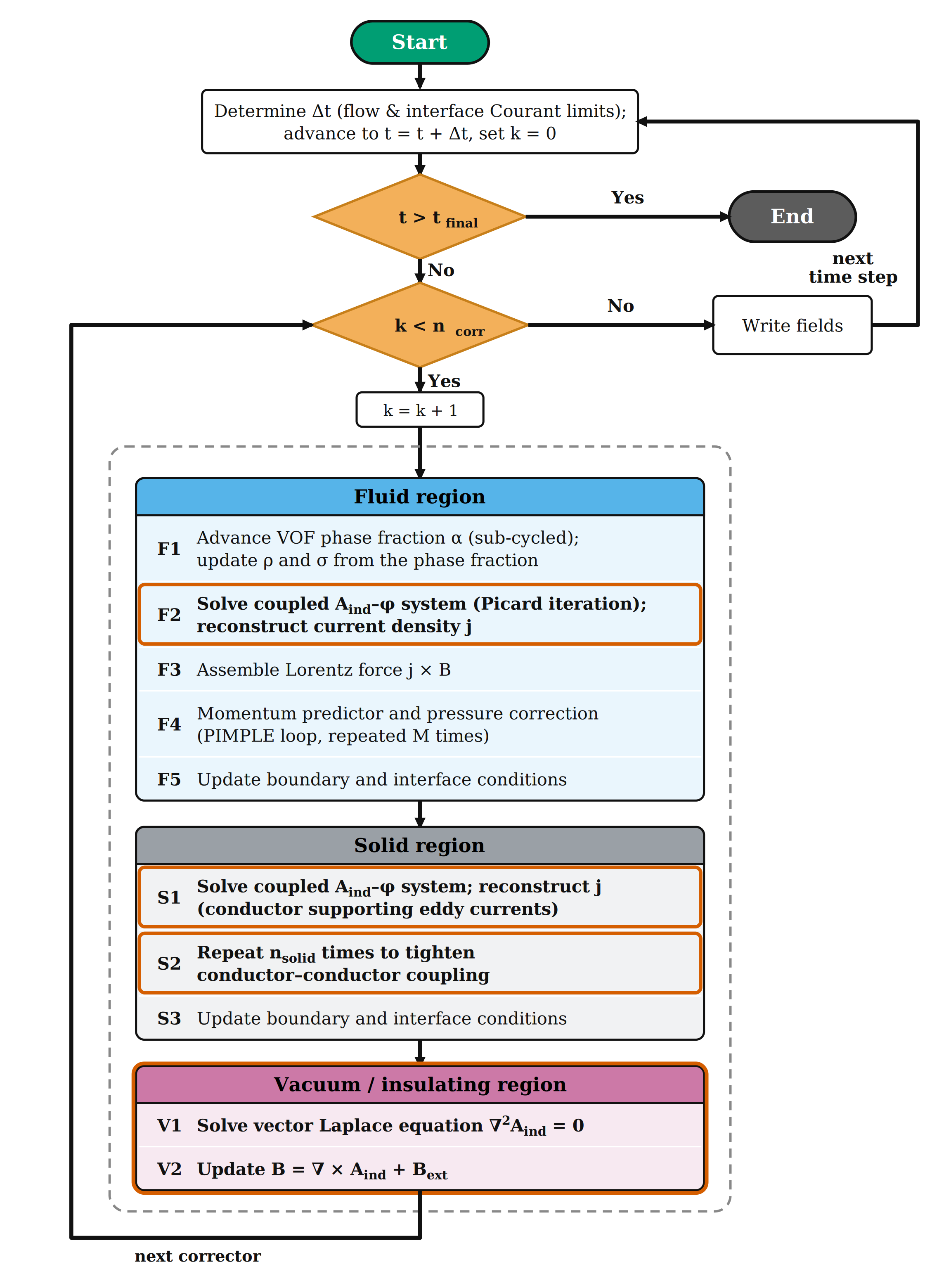}
  \caption{Flowchart of the per-time-step solution procedure. Steps highlighted with an orange outline and bold text denote the components introduced in this work.}
  \label{fig:flowchart}
\end{figure}

The complete per-time-step procedure assembles the region solvers of Secs.~\ref{free surface modeling}--\ref{boundary conditions} into a single segregated loop, summarized in Algorithm~\ref{alg:timestep} and depicted as a flowchart in Fig.~\ref{fig:flowchart}. At the start of each step the time-step size is updated to satisfy the flow and interface Courant limits. The step then advances through a fixed number of outer PIMPLE correctors, and within each corrector the regions are solved in turn, fluid first, then solid, then vacuum, so that every solver sees the most recent state of its neighbors through the interface conditions of Sec.~\ref{boundary conditions}.

For each fluid region the phase fraction is advanced by the sub-cycled VOF equation, the coupled $\boldsymbol{A}_{\mathrm{ind}}$--$\phi$ system is solved by the Picard iteration of Sec.~\ref{fluid mhd} and used to evaluate the current density and the Lorentz force, and the momentum and pressure equations are then solved in the usual PIMPLE sequence. For each solid region the scalar and vector potentials are solved by the procedure of Sec.~\ref{non fluid mhd}, repeated $n_{\mathrm{solid}}$ times per corrector to tighten the coupling between adjacent conductors. For each vacuum region the vector Laplace equation is solved and the magnetic field is updated. After the final corrector the fields are written and the solver advances to the next step.

{\renewcommand{\baselinestretch}{1}\selectfont
\begin{algorithm}[H]
\setlength{\algomargin}{2.5em}
\SetNlSkip{1em}
\SetInd{0.5em}{1.5em}
\caption{Time-step procedure}\label{alg:timestep}
Update the adaptive time step $\Delta t$ from the flow and interface Courant limits\;
\For{$k = 1$ \KwTo $n_{\mathrm{corr}}$}{
  \ForEach{fluid region}{
    advance the VOF phase fraction (sub-cycled)\;
    solve the coupled $\boldsymbol{A}_{\mathrm{ind}}$--$\phi$ system; evaluate $\boldsymbol{j}$, $\boldsymbol{j}\times\boldsymbol{B}$\;
    solve the momentum predictor and pressure-correction loop\;
  }
  \For{$m = 1$ \KwTo $n_{\mathrm{solid}}$}{
    \ForEach{solid region}{
      solve the coupled $\boldsymbol{A}_{\mathrm{ind}}$--$\phi$ system\;
    }
  }
  \ForEach{vacuum region}{
    solve $\nabla^2\boldsymbol{A}_{\mathrm{ind}}=0$; update $\boldsymbol{B}=\nabla\times\boldsymbol{A}_{\mathrm{ind}}+\boldsymbol{B}_{\mathrm{ext}}$\;
  }
}
Write fields and advance to the next time step\;
\end{algorithm}
}

The numerical parameters used for the calculations reported in Sec.~\ref{verification and validation} are collected in Table~\ref{num table}. Where a calculation is run in parallel, each region is decomposed independently by the \texttt{scotch} method, and the mapped interface patches of Sec.~\ref{boundary conditions} exchange data across rank boundaries. No under-relaxation is applied to any equation. The Picard loop of Sec.~\ref{fluid mhd} exits when the larger of the two initial residuals falls below $10^{-3}$, and its iteration ceiling is rarely reached; the table lists that ceiling alongside the count typically realized.

\begin{table}
\caption{\label{num table}Numerical parameters for the reported calculations. \emph{Duct} covers the six Shercliff and Hunt cases, \emph{Channel} and \emph{Transient duct} the two transient-field configurations, and \emph{Jet} the three mercury-jet cases. Typical counts are medians over the outer correctors. Where the two LMX-U calculations differ, entries are given as $0.3\,$T\,/\,$0.2\,$T. PCG, preconditioned conjugate gradient; PBiCG(Stab), (stabilized) preconditioned bi-conjugate gradient; GAMG, geometric-algebraic multigrid; DIC/DILU, (diagonal) incomplete Cholesky/LU.}
\begin{ruledtabular}
\setlength{\tabcolsep}{3pt}
\footnotesize
\begin{tabular}{lccccc}
Parameter & \shortstack{Duct\\{\scriptsize Sec.~\ref{sec:analytical_verification}}} & \shortstack{Channel\\{\scriptsize Sec.~\ref{sec:transient_verification}}} & \shortstack{Transient duct\\{\scriptsize Sec.~\ref{sec:transient_verification}}} & \shortstack{LMX-U\\{\scriptsize Sec.~\ref{sec:experimental_validation}}} & \shortstack{Jet\\{\scriptsize Sec.~\ref{sec:jet}}} \\[2pt]
\hline
Outer correctors $n_{\mathrm{corr}}$ & 3 & 10 & 3 & 1 & 3 \\
Picard iterations, maximum & 50 & 10 & 20 & 3 & 50 \\
Picard iterations, typical & 1--3 & 3 & 1 & 1 & 4 \\
Picard tolerance & \multicolumn{5}{c}{$10^{-3}$} \\
Solid sub-iterations $n_{\mathrm{solid}}$ & 3 & 0 & 0 & 1 & 3 \\
Pressure correctors & \multicolumn{5}{c}{3} \\
Non-orthogonal correctors & \multicolumn{5}{c}{0} \\
Under-relaxation & \multicolumn{5}{c}{none} \\
MPI ranks & serial & serial & 48 & 768\,/\,1536 & 48 \\
\hline
Solver for $\boldsymbol{A}_{\mathrm{ind}}$ & PCG/GAMG & \multicolumn{3}{c}{PBiCGStab/DIC} & PCG/GAMG \\
Solver for $\phi$ & \multicolumn{2}{c}{PCG/GAMG} & PCG/DIC & \multicolumn{2}{c}{PCG/GAMG} \\
Solver for $p_{\mathrm{rgh}}$ & \multicolumn{5}{c}{PCG/GAMG} \\
Solver for $\alpha$ & \multicolumn{5}{c}{PBiCG/DILU} \\
Solver for $\boldsymbol{u}$ & \multicolumn{5}{c}{PBiCG/DILU} \\
Tolerance, $\boldsymbol{A}_{\mathrm{ind}}$ & \multicolumn{5}{c}{$10^{-7}$} \\
Tolerance, $\phi$, $p_{\mathrm{rgh}}$ & $10^{-7}$ & $10^{-7}$ & $10^{-7}$ & $10^{-6}$ & $10^{-7}$ \\
Tolerance, $\alpha$ / $\boldsymbol{u}$ & \multicolumn{5}{c}{$10^{-8}$ / $10^{-6}$} \\
\end{tabular}
\end{ruledtabular}
\end{table}

\section{Verification and Validation}\label{verification and validation}

\subsection{Pressure-driven duct flows}
\label{sec:analytical_verification}

\begin{figure}[htbp]
  \centering
  \includegraphics[width=0.95\linewidth]{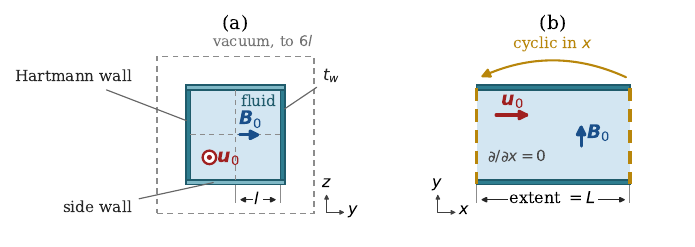}
  \caption{\label{fig:duct_geometry}Geometry of the Shercliff and Hunt duct benchmarks. (a)~Cross-section in the $y$--$z$ plane, to scale: the fluid (light blue) occupies $|y|,|z|\le l$, the Hartmann walls (dark teal) are normal to $\boldsymbol{B}_0$, which lies along $z$, and the side walls (medium teal) are parallel to it, both of thickness $t_w = 0.1\,l$. The bulk velocity $\boldsymbol{u}_0$ is directed in the $x$-direction; the dashed frame is the vacuum region, extending to $|y|,|z| = 6l$ and not drawn to scale. (b)~The same domain in the $x$--$y$ plane, cut through the Hartmann walls; the gold faces are a cyclic pair and the streamwise extent is $L = l$. Both configurations mesh their walls as solid regions of thickness $t_w$: the Shercliff walls form a single insulating region at $c = 10^{-13}$, whereas Hunt resolves its conducting Hartmann walls ($c = 10$) and its insulating side walls as two separate regions.}
\end{figure}

Two canonical MHD duct-flow configurations, the Shercliff case and the Hunt case, are employed to verify the code against known analytical solutions.~\cite{MullerBook} Both setups consider fully developed, pressure-driven flow in a square duct of half-width $l$ subject to a transverse external magnetic field $\boldsymbol{B}_0$ (the Hartmann direction). The walls perpendicular to $\boldsymbol{B}_0$ are called Hartmann walls; the remaining walls are the side walls. The duct geometry is illustrated in Fig.~\ref{fig:duct_geometry}. The two cases differ only in the electrical boundary condition imposed on the Hartmann walls, characterized by the wall conductance ratio $c \equiv \sigma_w t_w / (\sigma l)$, where $\sigma_w$ and $t_w$ are the wall electrical conductivity and thickness, respectively: Shercliff flow prescribes insulating Hartmann walls ($c \to 0$), while Hunt flow prescribes conducting ones. Side walls are electrically insulating in both cases. 

\subsubsection{Simulation parameters}

Because the analytical solutions describe fully developed flow, and are therefore independent of the streamwise coordinate, the computational domain is a streamwise-periodic slab rather than a long duct. The fluid core occupies $|y|,|z|\le l$ with $l = 0.1$, and the streamwise extent is $0.1$, resolved by ten cells with cyclic boundaries in $x$. The flow is driven by a uniform body force adjusted at each time step to hold the bulk velocity at $u_0 = 1$. This arrangement realizes the conditions of the analytical solution exactly and removes the entry length that a finite-length duct must otherwise accommodate; because it is inexpensive, it also permits a mesh to be tailored to each Hartmann number.

The remaining reference values are, in normalized units, dynamic viscosity $\mu = 1$, density $\rho = 10^{3}$, fluid electrical conductivity $\sigma = 10^{6}$, and magnetic permeability $\mu_0 = 10^{-8}$. In the Hunt case the Hartmann walls ($t_w = 0.01$, $\sigma_w = 10^{8}$) and the side walls ($\sigma_w = 10^{-6}$) are meshed as separate solid regions and solved as conductors; in the Shercliff case all four walls are insulating ($\sigma_w = 10^{-6}$) and are meshed as a single solid region, giving $c = 10^{-13}$, numerically indistinguishable from the insulating limit. A rectangular vacuum region surrounds the duct and extends to $|y|,|z| = 0.6$.

These choices give $Ha \equiv B_0 l\sqrt{\sigma/\mu} = 100\,B_0$, so that $B_0 = 0.1$, $1$ and $10$ produce the three cases $Ha = 10$, $100$ and $1000$ presented below; wall conductance ratios $c \to 0$ (Shercliff) and $c = \sigma_w t_w/(\sigma l) = 10$ (Hunt); and a magnetic Reynolds number $R_m = \mu_0\sigma u_0 l = 10^{-3}$.

A separate mesh is used at each Hartmann number, graded toward the walls so that the boundary layers are resolved equally well at all three. The first cell height at the Hartmann wall is $8.3\times10^{-4}$, $7.8\times10^{-5}$ and $7.9\times10^{-6}$ at $Ha = 10$, $100$ and $1000$, which places eight cells inside the Hartmann layer $\delta_H \equiv l/Ha$ and six inside the side layer $\delta_S \sim l/\sqrt{Ha}$ in every case, rather than allowing the resolution to degrade as $Ha$ increases. The meshes are tensor-product Cartesian, so the non-orthogonality vanishes identically and the skewness is below $10^{-12}$; the wall grading raises the maximum aspect ratio to $1.4\times10^{3}$ at $Ha = 1000$. All six duct calculations were run in serial, the meshes being small enough that domain decomposition is not worthwhile. Repeating the Hunt case at $Ha=100$ on $16$ ranks reproduces the serial result to the solver tolerance, the largest relative difference in any field being $5\times10^{-6}$ and the error norms of Table~\ref{err table} unchanged to the precision reported. These canonical cases lie in the low-$R_m$ range for which closed-form solutions are available; their purpose here is to verify that the solver reproduces the induced-field and current distributions correctly when the induced field is resolved rather than frozen, a quantity not solved for in the inductionless formulation.

\subsubsection{Shercliff flow}

In the Shercliff configuration all four walls are electrically insulating. The induced currents are therefore confined entirely to the fluid, completing their circuit within the conducting medium. This produces two distinct boundary-layer structures: thin Hartmann layers of thickness $\delta_H = l/Ha$ adjacent to the Hartmann walls, and thicker side-wall layers of thickness $\delta_S \sim l/\sqrt{Ha}$ adjacent to the side walls. As $Ha$ increases, both layers sharpen, and the core velocity approaches a flat, nearly uniform profile. The induced magnetic field $B_{\mathrm{ind}}$ is carried entirely by the fluid current and exhibits a nearly linear variation across the duct in the Hartmann direction, vanishing at the insulating walls.

Figs.~\ref{shercliff U} and~\ref{shercliff B} compare simulation results with the analytical solutions of Ref.~\onlinecite{MullerBook} for $Ha = 10$, $100$, and $1000$. The velocity profiles in Fig.~\ref{shercliff U} show excellent agreement in both the Hartmann (panel~a) and side-wall (panel~b) directions across all three Hartmann numbers, with the characteristic flattening of the core and the progressive thinning of the boundary layers as $Ha$ increases. The induced-field distributions in Fig.~\ref{shercliff B} likewise agree closely with the analytical solution at all Hartmann numbers and at both the duct midplane ($z/l = 0$) and the off-center slice ($z/l = 0.75$).

\subsubsection{Hunt flow}

In the Hunt configuration the Hartmann walls are conducting, at the finite conductance ratio $c=10$. Return currents now preferentially flow through the low-resistance Hartmann walls rather than through the fluid interior, drastically altering the current topology. The most striking consequence is the formation of high-velocity jets near the side walls: current re-enters the fluid from the Hartmann walls at the side-wall region, generating a strong Lorentz force that accelerates the fluid there. These side-wall jets, which exceed the core velocity at high $Ha$, are a hallmark of the Hunt flow and are absent in the Shercliff case. The induced magnetic field is concentrated near the fluid--Hartmann-wall interface and exhibits a more complex spatial structure than in the Shercliff case.

Figs.~\ref{hunt U} and~\ref{hunt B} show the corresponding Hunt-case results. The Hartmann-direction profiles (Fig.~\ref{hunt U}a) are again in close agreement with the analytical solution and share the same qualitative character as the Shercliff case. The side-wall profiles (Fig.~\ref{hunt U}b), however, are qualitatively different: the simulation correctly reproduces the pronounced velocity jets near $z/l \to 1$ at high $Ha$ and the transition from a parabolic to a jet-dominated profile as $Ha$ increases. The induced-field comparisons in Fig.~\ref{hunt B} confirm good agreement with the analytical solution for all three Hartmann numbers.

\begin{table}
\caption{\label{err table}Relative $L_2$ and $L_\infty$ errors against the analytical solutions of Ref.~\onlinecite{MullerBook}, in percent, over the full duct cross-section. Fields are normalized by their maximum absolute value before the comparison and both norms scaled by the peak analytical magnitude. The induced field is evaluated from an unlimited least-squares gradient (see text). (Sec.~\ref{sec:analytical_verification})}
\begin{ruledtabular}
\begin{tabular}{lcccc}
 & \multicolumn{2}{c}{$u_x$ (cross-section)} & \multicolumn{2}{c}{$b_x$ (cross-section)} \\
Case ($Ha$) & $L_2$ (\%) & $L_\infty$ (\%) & $L_2$ (\%) & $L_\infty$ (\%) \\
\hline
Shercliff (10)   & 0.05 & 0.13 & 0.31 & 1.24 \\
Shercliff (100)  & 0.10 & 0.19 & 0.38 & 2.52 \\
Shercliff (1000) & 0.26 & 0.55 & 2.23 & 5.42 \\
Hunt (10)        & 0.29 & 0.82 & 0.44 & 2.17 \\
Hunt (100)       & 0.23 & 0.77 & 1.51 & 3.70 \\
Hunt (1000)      & 0.16 & 0.77 & 1.70 & 4.12 \\
\end{tabular}
\end{ruledtabular}
\end{table}

\subsubsection{Accuracy and mesh sensitivity}

\begin{figure}[htbp]
  \centering
  \includegraphics[width=\linewidth]{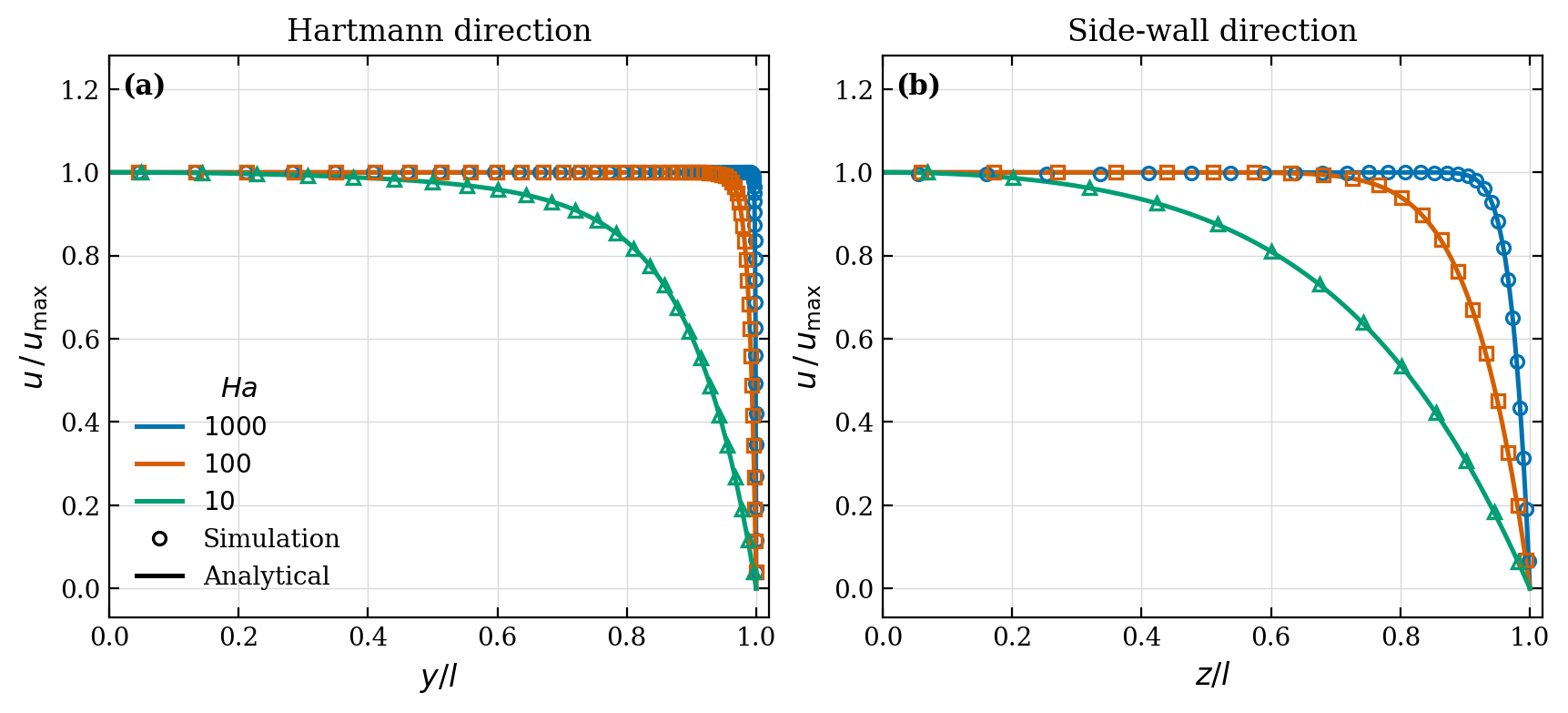}
  \caption{%
    Normalized velocity profiles for the Shercliff flow at $Ha = 10$, $100$, and $1000$. (a)~Profile along the Hartmann direction ($y/l$) at the duct midplane. (b)~Profile along the side-wall direction ($z/l$) at the duct midplane. Solid lines denote analytical solutions,\cite{MullerBook} whereas markers indicate simulation results.%
  }
  \label{shercliff U}
\end{figure}

\begin{figure}[htbp]
  \centering
  \includegraphics[width=\linewidth]{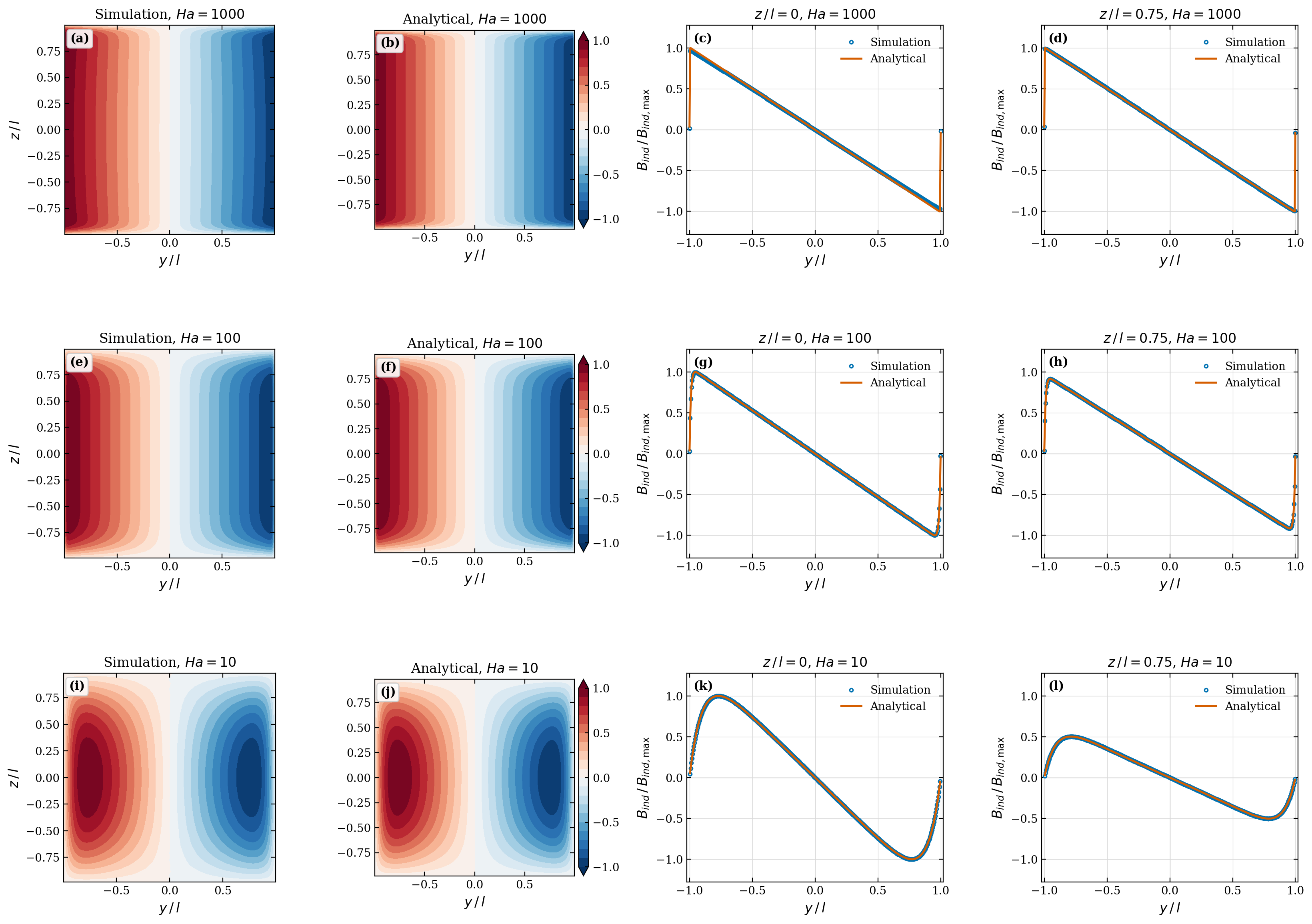}
  \caption{%
    Normalized induced magnetic field $B_{\mathrm{ind}}/B_{{\mathrm{ind}},\max}$ (where $\boldsymbol{B}_{\mathrm{ind}} = \nabla\times\boldsymbol{A}_{\mathrm{ind}}$) for the Shercliff flow at $Ha = 10$, $100$, and $1000$. Panels (a,\,b), (e,\,f), and (i,\,j) show two-dimensional contour maps of the simulation and analytical solutions, respectively. Panels (c,\,g,\,k) compare line profiles at the duct midplane ($z/l = 0$), and panels (d,\,h,\,l) at the off-center slice ($z/l = 0.75$). Solid lines: analytical solution;~\cite{MullerBook} circles: simulation.%
  }
  \label{shercliff B}
\end{figure}

\begin{figure}[htbp]
  \centering
  \includegraphics[width=\linewidth]{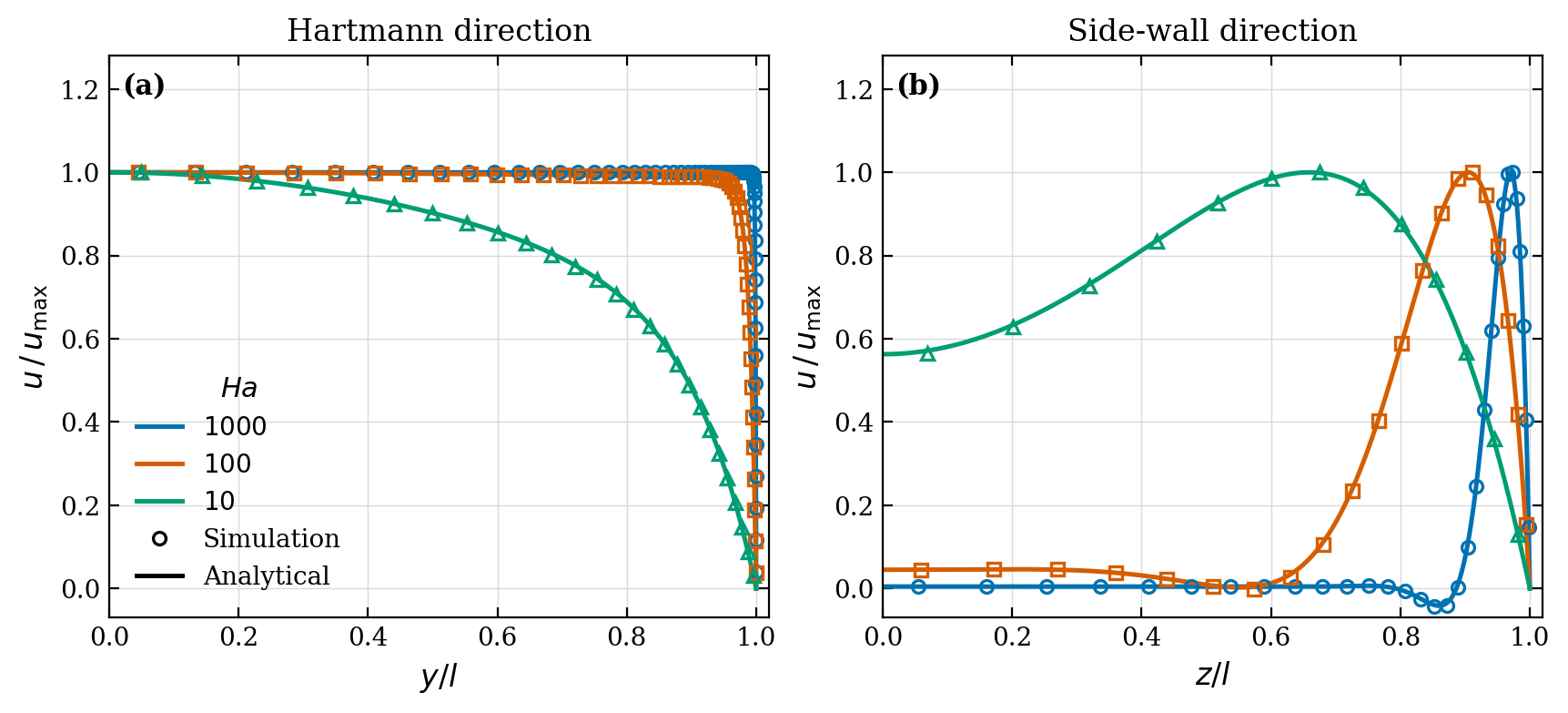}
  \caption{Normalized velocity profiles for the Hunt flow at $Ha = 10$, $100$, and $1000$. Layout identical to Fig.~\ref{shercliff U}.}
  \label{hunt U}
\end{figure}

\begin{figure}[htbp]
  \centering
  \includegraphics[width=\linewidth]{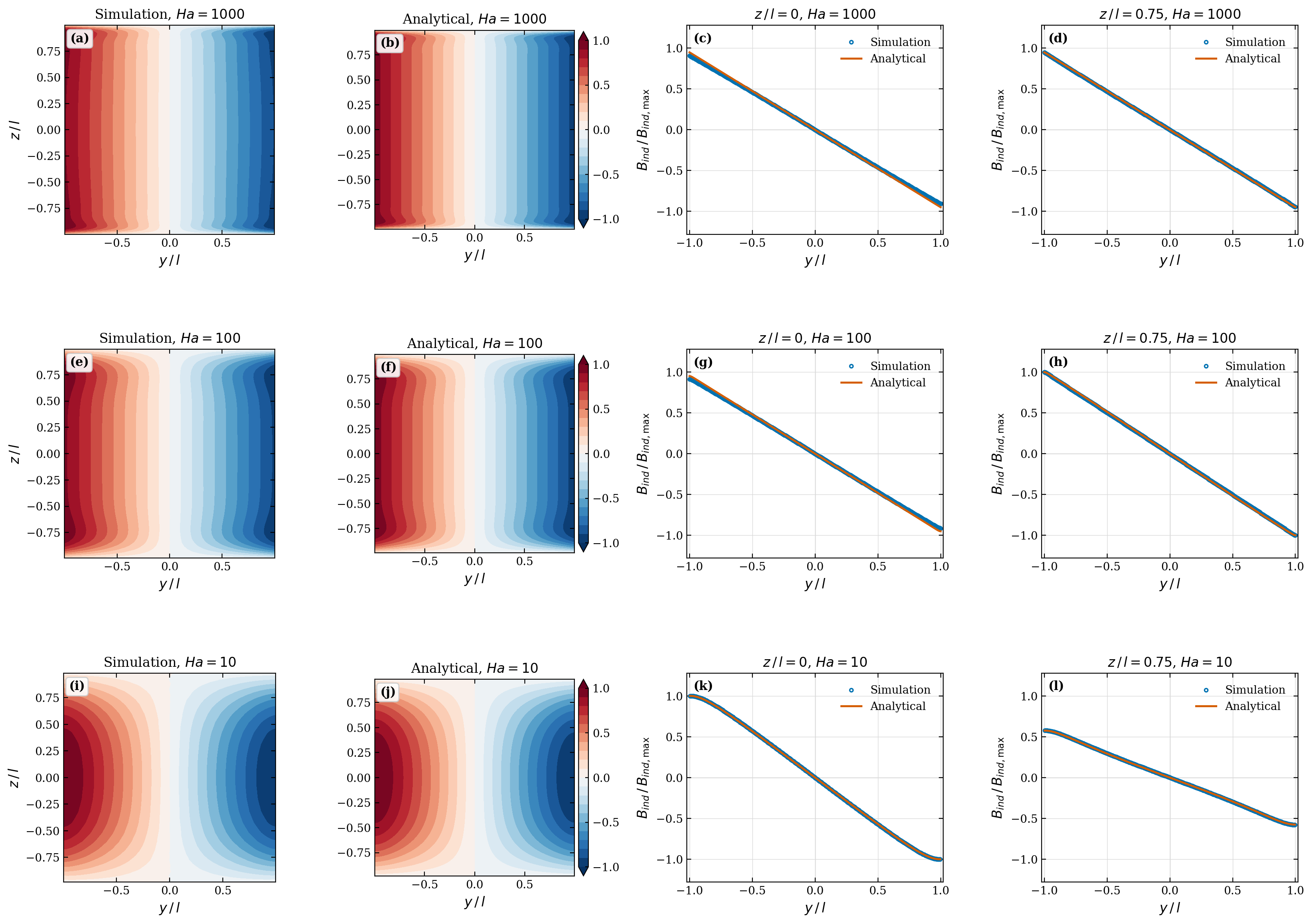}
  \caption{%
    Normalized induced magnetic field $B_{\mathrm{ind}}/B_{{\mathrm{ind}},\max}$ for the Hunt flow at $Ha = 10$, $100$, and $1000$. Layout identical to Fig.~\ref{shercliff B}.%
  }
  \label{hunt B}
\end{figure}

The agreement is quantified in Table~\ref{err table}. Both the streamwise velocity and the induced field are compared over the entire cross-section rather than along the two symmetry lines only, so that the two quantities are assessed on the same footing; the analytical velocity is evaluated at arbitrary points in the plane from the same series expansion that supplies the profiles of Figs.~\ref{shercliff U} and~\ref{hunt U}. Computed and analytical fields are each normalized by their own maximum, as in the figures, so that the norms measure profile shape, and both norms are scaled by the peak magnitude of the analytical field, so that $L_2 \le L_\infty$ by construction.

The induced field is a derived quantity, $b_x = (\nabla\times\boldsymbol{A}_{\mathrm{ind}})_x$, and its evaluation deserves comment. The solver's default cell-limited gradient, appropriate for the transported fields, clips the sharp gradient of $\boldsymbol{A}_{\mathrm{ind}}$ in the near-wall cells and yields spurious pointwise extrema in $b_x$ that reach tens of percent at $Ha = 1000$. The values reported here are obtained from an unlimited least-squares gradient. This is a post-processing choice rather than a property of the solution, but it is the difference between a diagnostic that converges under refinement and one that does not, and we record it so that the comparison can be reproduced.

Across all six cases the velocity errors remain below $0.3\%$ in $L_2$ and $0.9\%$ in $L_\infty$, and the induced-field errors below $2.3\%$ and $5.5\%$ respectively. The largest errors occur at $Ha = 1000$, where the Hartmann layer is thinnest, and the pointwise maxima are attained inside the Hartmann and side layers rather than in the core.

To establish the observed order of accuracy, and to assess how sensitive the induced field is to resolution, a three-level refinement family was run at $Ha=100$ for the Shercliff case. The middle level is the mesh of Table~\ref{err table}, so the family is anchored to the results already reported, and the coarse and fine levels bracket it at a cross-plane refinement ratio of approximately $1.5$, giving $19\,040$, $42\,840$ and $97\,280$ fluid cells. Only the cross-plane is refined: the analytical solution is independent of the streamwise coordinate, so the streamwise discretization contributes no error to the comparison, and the order reported here is the cross-plane order with representative cell size $h = N_\perp^{-1/2}$, where $N_\perp$ is the number of cells in the cross-section.

Table~\ref{order table} gives the errors on each level together with the observed order between successive levels, $p = \ln(e_{\mathrm{coarse}}/e_{\mathrm{fine}})/\ln(h_{\mathrm{coarse}}/h_{\mathrm{fine}})$. The streamwise velocity converges at second order in both norms, the two estimates agreeing to within $0.1$. The induced field converges at approximately first order, which is the expected behaviour for a quantity obtained by differentiating the computed vector potential, one order being lost to the curl.

\begin{table}
\caption{\label{order table}Mesh-refinement study of the Shercliff case at $Ha=100$. Level L2 is the mesh of Table~\ref{err table}, $N_\perp$ the number of cross-section cells, and the errors are defined as there; $p$ is the observed order between successive levels. (Sec.~\ref{sec:analytical_verification})}
\begin{ruledtabular}
\begin{tabular}{lcccc}
 & \multicolumn{2}{c}{$u_x$} & \multicolumn{2}{c}{$b_x$} \\
Level ($N_\perp$) & $L_2$ (\%) & $L_\infty$ (\%) & $L_2$ (\%) & $L_\infty$ (\%) \\
\hline
L1 ($1904$) & 0.219 & 0.422 & 0.587 & 3.086 \\
L2 ($4284$) & 0.101 & 0.194 & 0.381 & 2.524 \\
L3 ($9728$) & 0.044 & 0.085 & 0.217 & 1.752 \\
\hline
\multicolumn{5}{c}{Observed order $p$} \\
L1$\to$L2 & 1.91 & 1.92 & 1.07 & 0.50 \\
L2$\to$L3 & 2.01 & 2.02 & 1.37 & 0.89 \\
\end{tabular}
\end{ruledtabular}
\end{table}

Overall, the simulation results are in good agreement with the analytical solutions~\cite{MullerBook} for both test cases across the full range of Hartmann numbers, providing confidence in the correctness of the electromagnetic and momentum-transport implementations.

\subsection{Channel and duct under a transient applied field}\label{sec:transient_verification}

The duct flows of Sec.~\ref{sec:analytical_verification} are driven by a pressure gradient under a static applied field. They verify that the induced field is computed correctly, but they do not exercise the situation for which resolving the induced field is indispensable, namely a flow driven by a time-varying applied field. This section adds that verification, using the liquid metal channel flow under a transient magnetic field formulated by Smolentsev\cite{SmolentsevFED25} and recently recreated by Endeve~\textit{et al.}\cite{EndeveFED26}

The configuration models a breeding-blanket flow channel during the current quench of a plasma disruption. A liquid metal flows along a channel of half-width $a$ under a transverse magnetic field $B_{z0}$ that is constant in time, while the field component aligned with the flow decreases linearly, $\partial B_{x0}/\partial t=\gamma$. The decaying component induces an electromotive force that drives current loops closing in the channel cross-section; their interaction with the constant transverse field produces a streamwise Lorentz force that is antisymmetric about the channel centerline, and hence a velocity profile with positive and negative lobes carrying \emph{zero net flow rate}. This flow is generated entirely by $\partial\boldsymbol{B}/\partial t$; it has no counterpart in the inductionless formulation, in which the induced field is frozen and the electromotive force from its time derivative is absent altogether.

The decaying component is imposed through its vector potential, $\partial\boldsymbol{A}_{\mathrm{ext}}/\partial t$ being prescribed in closed form as described in Sec.~\ref{fluid mhd}. For a spatially uniform applied field changing at a rate $\dot{\boldsymbol{B}}_{\mathrm{ext}}$ the symmetric gauge
\begin{equation}\label{Aext gauge}
\frac{\partial\boldsymbol{A}_{\mathrm{ext}}}{\partial t}
= \tfrac{1}{2}\,\dot{\boldsymbol{B}}_{\mathrm{ext}}\times(\boldsymbol{x}-\boldsymbol{x}_{0})
\end{equation}
is used for the three-dimensional duct, with $\boldsymbol{x}_{0}$ a prescribed reference point. It is not suitable for the one-dimensional channel, where it would place half of the transient electromotive force in a component varying along the collapsed coordinate, so that half of the drive is lost; there a gauge independent of that coordinate is used instead,
\begin{equation}\label{Aext landau}
\frac{\partial\boldsymbol{A}_{\mathrm{ext}}}{\partial t}
= \tfrac{1}{2}\bigl(\dot{\boldsymbol{B}}_{\mathrm{ext}}\!\cdot\hat{\boldsymbol{e}}\bigr)\,
\hat{\boldsymbol{e}}\times(\boldsymbol{x}-\boldsymbol{x}_{0})
+ \bigl[\bigl(\hat{\boldsymbol{e}}\times\dot{\boldsymbol{B}}_{\mathrm{ext}}\bigr)\!\cdot(\boldsymbol{x}-\boldsymbol{x}_{0})\bigr]\hat{\boldsymbol{e}},
\end{equation}
with $\hat{\boldsymbol{e}}$ the collapsed direction. Both gauges have curl $\dot{\boldsymbol{B}}_{\mathrm{ext}}$ and so represent the same applied field.

Three cases are reproduced from the two configurations of Ref.~\onlinecite{SmolentsevFED25}: the one-dimensional channel of its Sec.~3.2, and the square duct of its Sec.~3.1 in a steady and then a transient state.

\subsubsection{One-dimensional channel}

\begin{figure}[htbp]
  \centering
  \includegraphics[width=\linewidth]{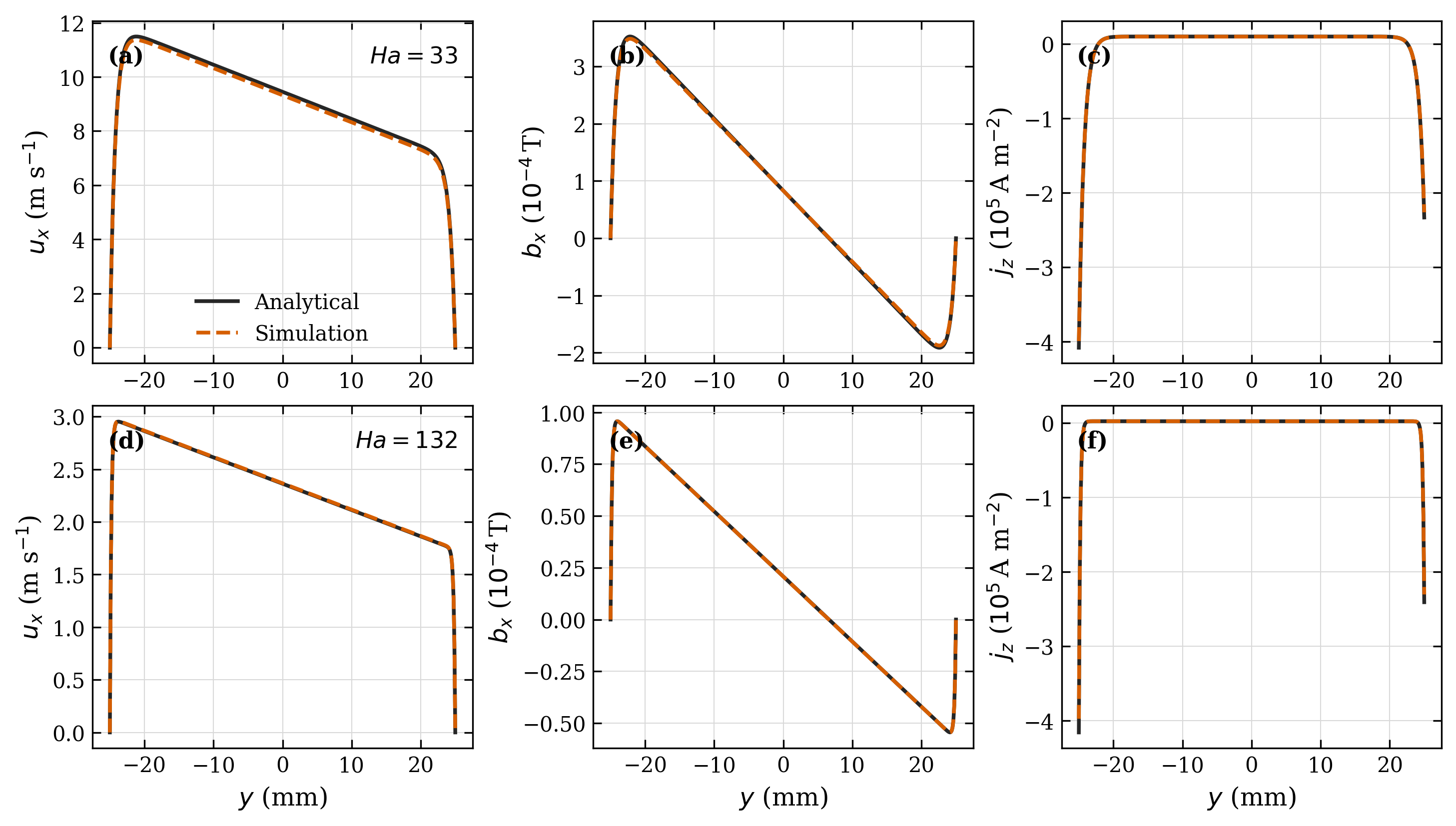}
  \caption{\label{transient channel}Liquid metal channel flow under a transient applied magnetic field, at $Ha=33$ (a--c) and $Ha=132$ (d--f), compared with the closed-form solution of Ref.~\onlinecite{SmolentsevFED25}. (a,\,d)~Streamwise velocity; (b,\,e)~induced magnetic field; (c,\,f)~current density. Solid lines denote the analytical solution and dashed lines the simulation.}
\end{figure}

The channel of Ref.~\onlinecite{SmolentsevFED25}, Sec.~3.2, admits a closed-form solution: eutectic lead--lithium with $\sigma=0.7\times10^{6}\,\mathrm{S\,m^{-1}}$ and $\mu=10^{-3}\,\mathrm{Pa\,s}$, half-width $a=25\,\mathrm{mm}$, pressure gradient $G=-500\,\mathrm{Pa\,m^{-1}}$, field decay rate $\gamma=-5\,\mathrm{T\,s^{-1}}$, insulating walls, at transverse fields of $0.05$ and $0.2\,\mathrm{T}$ corresponding to $Ha=33$ and $132$. The cross-channel direction is resolved with $640$ cells graded toward the walls, the first at $17\,\mu\mathrm{m}$, which places $22$ cells inside the Hartmann layer at $Ha=33$ and eight at $Ha=132$. Figure~\ref{transient channel} compares the computed streamwise velocity, induced field and current density with that solution. The agreement is close: the relative $L_2$ errors are $1.33\%$, $1.27\%$ and $1.70\%$ in $u_x$, $b_x$ and $j_z$ at $Ha=33$, and $0.17\%$, $0.10\%$ and $3.76\%$ at $Ha=132$, the largest value arising from the current density peak within the Hartmann layer. The antisymmetric velocity profile and the two counter-directed current sheets adjacent to the Hartmann walls are reproduced in both cases.

\subsubsection{Square duct}

This configuration extends the same transient-field problem to three dimensions: the square duct of Ref.~\onlinecite{SmolentsevFED25}, Sec.~3.1, with $a=h=0.1\,\mathrm{m}$ and insulating walls. This duct is not one of the pressure-driven cases of Sec.~\ref{sec:analytical_verification}; the applied field varies in time here as well, and only the geometry is shared. The reference tabulates the dimensionless flow rate $\tilde Q$, and because the transient-driven part of the solution carries no net flow, $\tilde Q$ measures the pressure-driven component and provides an independent check. At $Ha=500$ the computed value converges to the tabulated one under mesh refinement: a $100\times200$ cross-plane mesh, which spans the Hartmann layer $\delta_H=a/Ha$ with two to three cells, gives $\tilde Q=7.799\times10^{-3}$, exceeding the reference value of $7.680\times10^{-3}$ by $1.55\%$, whereas a $100\times400$ mesh with $16$ cells inside $\delta_H$ gives $7.684\times10^{-3}$ and recovers the four-significant-figure reference value to $0.05\%$.

The transient evolution of that same duct is reported next, for which both references report time histories. The fluid starts from rest and is driven simultaneously by a pressure gradient $G=-50\,\mathrm{Pa\,m^{-1}}$ and by a streamwise field decaying at $\gamma=-100\,\mathrm{T\,s^{-1}}$, giving $Sm=(\gamma/G)\sqrt{\sigma\nu\rho}=52.9$. It is repeated at four constant transverse fields from $B_{z0}=0.2$ to $4.0\,\mathrm{T}$, spanning $Ha=529$ to $10\,583$, on a $200\times800$ graded cross-plane mesh that resolves the Hartmann layer with at least nine cells at every field.

The formulation of Ref.~\onlinecite{SmolentsevFED25} is two-dimensional and fully developed, its velocity purely streamwise, and the present calculation adopts the same restriction. The decaying component therefore acts through its rate alone, $\partial\boldsymbol{A}_{\mathrm{ext}}/\partial t$, while its instantaneous value $B_{x0}=\gamma t$ is omitted from the total field of Eq.~(\ref{B reconstruction}); retaining it would add a cross-plane Lorentz force $\boldsymbol{j}\times B_{x0}\hat{\boldsymbol{x}}$, and with it a secondary circulation that the reference does not represent.

\begin{table}[htbp]
\caption{\label{transient b table}Peak induced field $b_{x,\max}$ in the transient square duct, taken over the fluid region, against Ref.~\onlinecite{SmolentsevFED25}. (Sec.~\ref{sec:transient_verification})}
\begin{ruledtabular}
\begin{tabular}{lccc}
$B_{z0}$ (T) & This work (T) & Ref.~\onlinecite{SmolentsevFED25} (T) & Difference \\
\hline
$0.2$ & $0.2556$ & $0.2559$ & $-0.12\%$ \\
$0.5$ & $0.2458$ & $0.2461$ & $-0.11\%$ \\
$1.0$ & $0.2259$ & $0.2264$ & $-0.23\%$ \\
$4.0$ & $0.1400$ & $0.1410$ & $-0.72\%$ \\
\end{tabular}
\end{ruledtabular}
\end{table}

Table~\ref{transient b table} collects the peak induced field, which agrees with Ref.~\onlinecite{SmolentsevFED25} to within $0.72\%$ at every field strength. Measured against the applied component, $b_{x,\max}/B_{z0}$ falls from $1.28$ at $B_{z0}=0.2\,\mathrm{T}$ to $0.035$ at $4.0\,\mathrm{T}$. The first of these is the substantive point: the induced field there exceeds the constant applied component, so the total field in the duct is dominated by the induced contribution rather than perturbed by it. That regime is reached not through advective stretching, the magnetic Reynolds number formed with the instantaneous peak velocity being only $0.086$, but because the transient electromotive force $-\partial\boldsymbol{A}/\partial t$ generates the induced field directly. An inductionless treatment does not contain that term and returns no induced field at all in this configuration, at any $R_m$.

\begin{figure}[htbp]
  \centering
  \includegraphics[width=\linewidth]{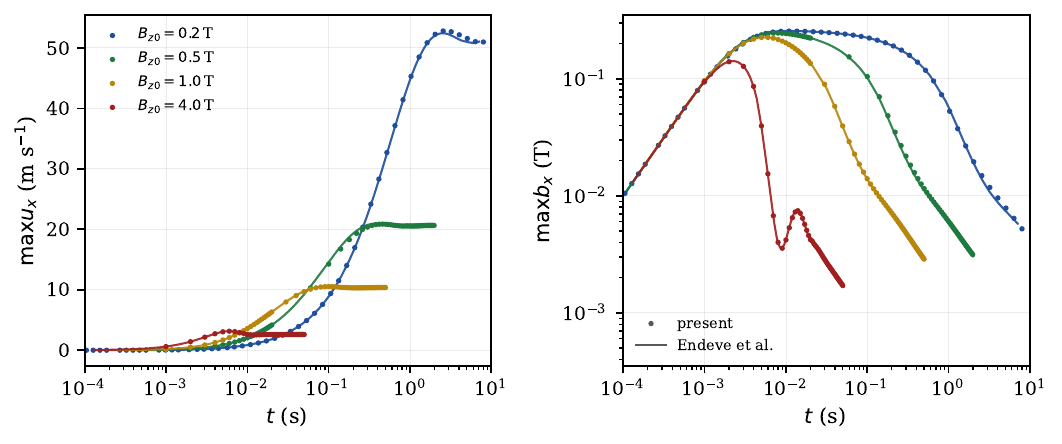}
  \caption{\label{transient duct evolution}Maximum streamwise velocity (left) and maximum induced field (right) versus time in the transient square duct, at $B_{z0}=0.2$, $0.5$, $1.0$ and $4.0\,\mathrm{T}$. Markers are the present calculation, taken over the discrete solution in the fluid region; solid lines are Ref.~\onlinecite{EndeveFED26}. The time axis is logarithmic and the $0.2\,\mathrm{T}$ history is the finest-timestep calculation.}
\end{figure}

Figure~\ref{transient duct evolution} compares the full time histories of the maximum streamwise velocity and the maximum induced field against Ref.~\onlinecite{EndeveFED26}, each taken over the discrete solution in the fluid region. The induced field at $4\,\mathrm{T}$ passes through a minimum and recovers, a feature Ref.~\onlinecite{SmolentsevFED25} attributes to the location of the velocity maximum migrating from the duct corner to the centerline. The present calculation reproduces it, which is a sensitive check because it depends on where in the cross-section the Lorentz force is applied rather than on its magnitude. We place both the minimum and the subsequent recovery within $0.8\%$ of Ref.~\onlinecite{EndeveFED26} in magnitude and within $1.5\%$ in time. The two references differ here, Ref.~\onlinecite{SmolentsevFED25} placing the feature earlier and an order of magnitude stronger; Ref.~\onlinecite{EndeveFED26} reports the same deviation and attributes it to the two-dimensional model of Ref.~\onlinecite{SmolentsevFED25}, and the present calculation follows it. The timestep is resolved against the magnetic diffusion time $\tau_m=\mu_0\sigma L^2=8.8\times10^{-3}\,\mathrm{s}$, which is the same at every field strength: the $0.2\,\mathrm{T}$ case was repeated from rest at three timesteps in the ratio four, with the mesh, the discretization and every physical parameter held fixed, showing monotone convergence in both quantities.
\begin{table}[htbp]
\caption{\label{transient u table}Maximum streamwise velocity in the transient square duct against Ref.~\onlinecite{EndeveFED26}, at every sampled time for which it exceeds $0.2\,\mathrm{m\,s^{-1}}$, taken over the discrete solution in the fluid region. (Sec.~\ref{sec:transient_verification})}
\begin{ruledtabular}
\begin{tabular}{lcccc}
$B_{z0}$ (T) & $Ha$ & $u_{x,\max}$ (m\,s$^{-1}$) & Mean difference & Range \\
\hline
$0.2$ & $529$     & $52.07$ & $+0.7\%$ & $-0.8$ to $+1.6\%$ \\
$0.5$ & $1323$    & $20.86$ & $-0.3\%$ & $-2.4$ to $+0.4\%$ \\
$1.0$ & $2646$    & $10.53$ & $-0.0\%$ & $-0.5$ to $+0.5\%$ \\
$4.0$ & $10\,583$ & $3.145$ & $+0.2\%$ & $-0.7$ to $+1.5\%$ \\
\end{tabular}
\end{ruledtabular}
\end{table}

On that basis the agreement is close throughout. The maximum velocity differs from Ref.~\onlinecite{EndeveFED26} by less than $0.8\%$ in the mean at every field strength and by at most $2.4\%$ at every sampled time in all four histories (Table~\ref{transient u table}); at $4\,\mathrm{T}$, where the transient completes within the simulated window, the peak velocity and the time at which it occurs agree to $0.3\%$. The maximum induced field agrees to better than $1.7\%$ until it has decayed to a few percent of its peak, beyond which the two differ by about $6\%$.

Two features of this case bear on the regime the solver is intended for. The induced field exceeds the applied component while the fluid is still nearly at rest, at $R_m=0.086$; and the calculations then run to their terminal state, the $0.2\,\mathrm{T}$ case passing through $R_m=4.6$. A single calculation therefore carries the solver from $R_m$ below $0.1$ to $R_m$ above unity, agreeing with Ref.~\onlinecite{EndeveFED26} throughout and recovering the closed-form terminal profile of Ref.~\onlinecite{SmolentsevFED25} to within $1\%$ in peak velocity.

\begin{figure}[htbp]
  \centering
  \includegraphics[width=\linewidth]{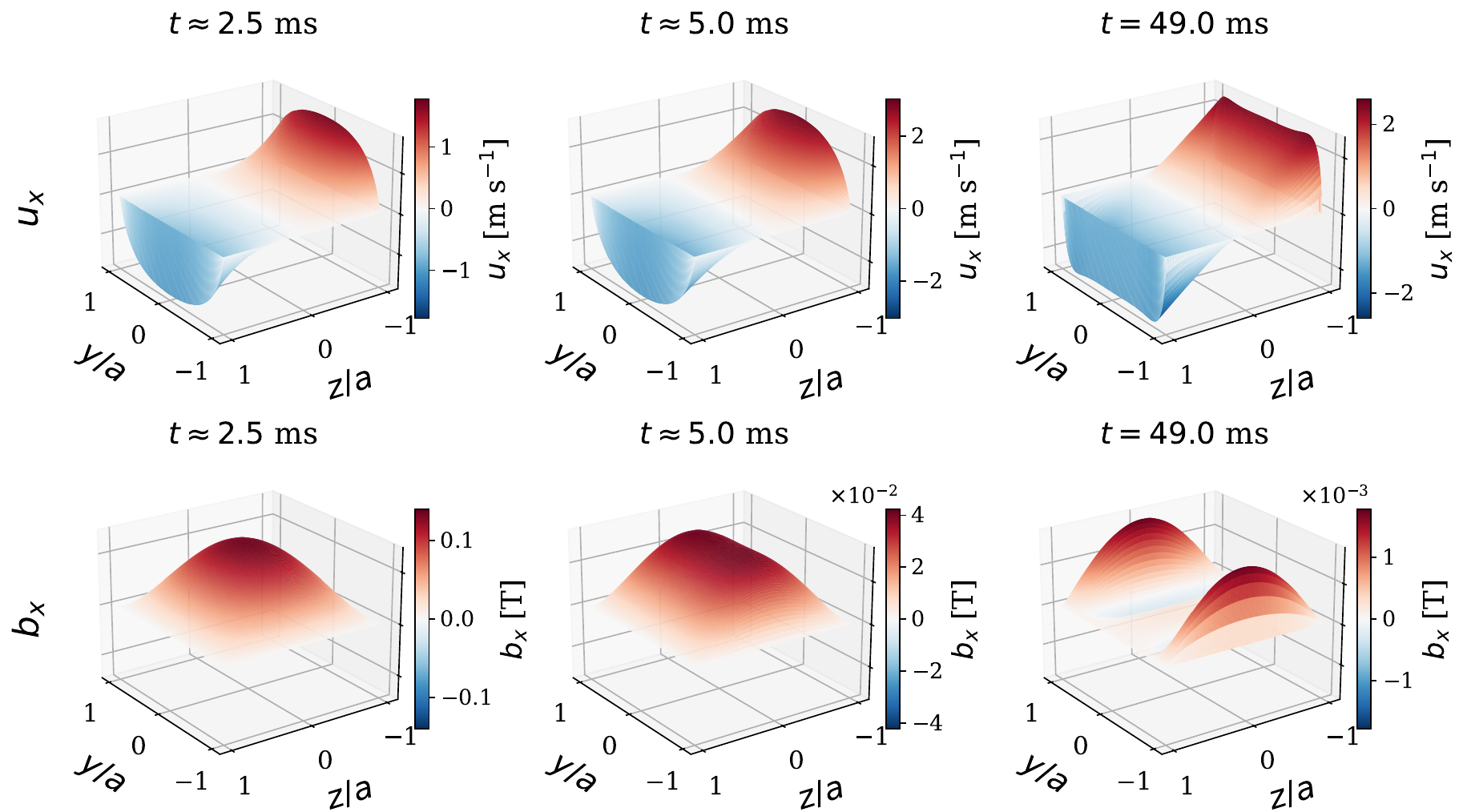}
  \caption{\label{transient duct xsec}Streamwise velocity $u_x$ (top) and induced field $b_x$ (bottom) over the duct cross-section at $B_{z0}=4\,\mathrm{T}$, drawn as surfaces at $t\approx2.5$, $5.0$ and $49.0\,\mathrm{ms}$. The presentation and the snapshot times follow Fig.~12 of Ref.~\onlinecite{EndeveFED26}, whose $v_z$ and $b_z$ are the $u_x$ and $b_x$ here.}
\end{figure}

Figure~\ref{transient duct xsec} shows the cross-sectional velocity and induced field at $4\,\mathrm{T}$ as they develop, in the presentation used by Ref.~\onlinecite{EndeveFED26}. The velocity is the antisymmetric pair of counter-directed lobes anticipated above, with the extrema at midspan in the side-wall direction and within the Hartmann layer. The induced field is drawn in the bottom row, each panel on its own scale. Its amplitude falls by nearly two orders of magnitude across the three times while the single dome on the duct axis gives way to a pair of ridges, and the minimum and recovery discussed above fall within that interval.

These three cases establish that the solver reproduces the flow driven by a time-varying applied field, quantitatively and in configurations drawn from the literature. None involves a free surface; no reference solution or experimental data exists for a free-surface flow under a transient field, which is why the verification is performed in the closed geometries for which the reference provides one.

\subsection{Experimental validation: the LMX-U channel}
\label{sec:experimental_validation}
 
\begin{figure}
\includegraphics[width=\linewidth]{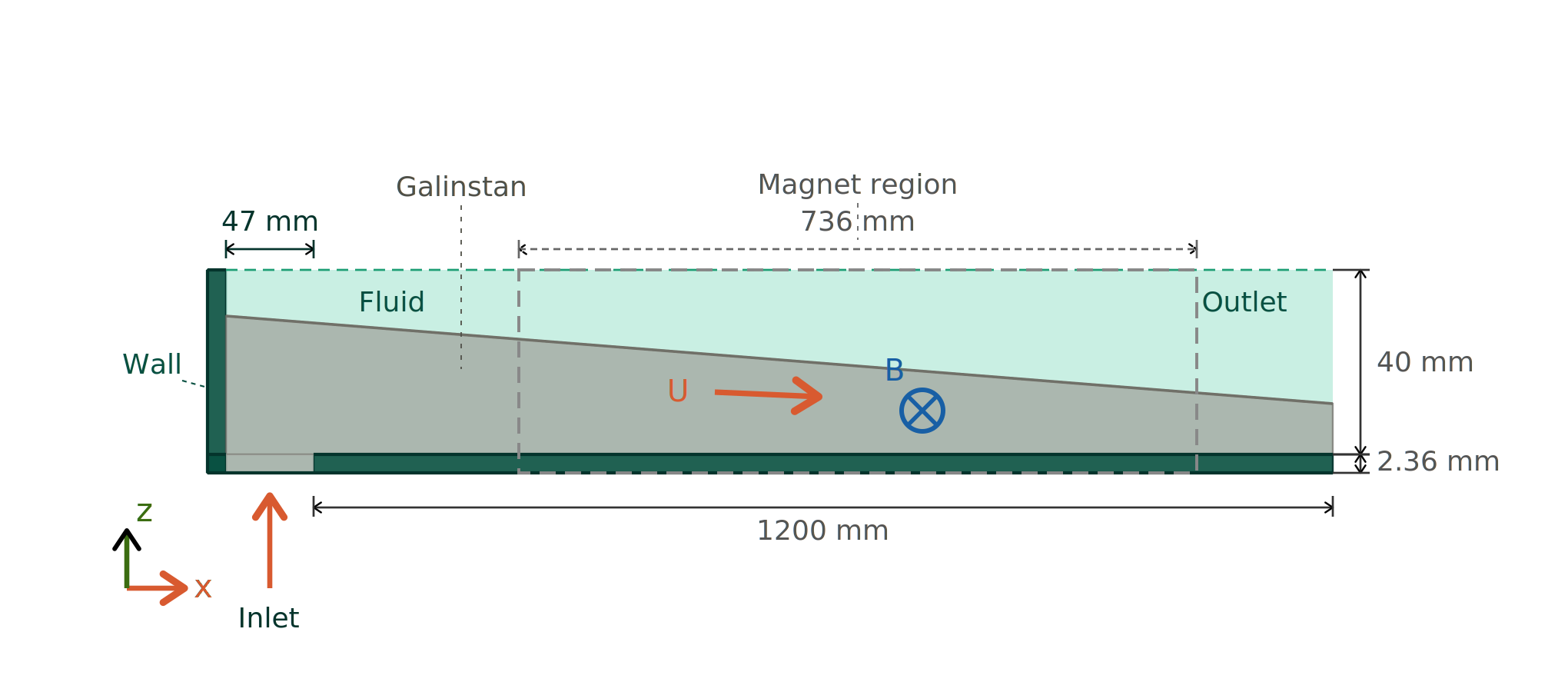}
\caption{Schematic of the LMX-U facility, side view in the $x$--$z$ plane. The channel is 1200~mm long, 105~mm wide and 40~mm high, with 2.36~mm walls. Galinstan (gray) enters through the 47~mm inlet region and exits freely at the outlet. The dashed rectangle marks the 736~mm electromagnet, which applies $\boldsymbol{B}$ along $y$, into the page; the flow $\boldsymbol{u}$ is in $+x$.}
\label{LMX blueprint}
\end{figure}

The solver is validated against free-surface height measurements obtained from the LMX-U experiment,\cite{SunNF23} an open-channel liquid metal flow facility located at the Princeton Plasma Physics Laboratory (PPPL). A schematic of the device is shown in Fig.~\ref{LMX blueprint}. The channel has a length of $1200\,\mathrm{mm}$, a width of $105\,\mathrm{mm}$, and a height of $40\,\mathrm{mm}$, with a wall thickness of $2.36\,\mathrm{mm}$. An electromagnet spanning $736\,\mathrm{mm}$ of the channel applies a transverse magnetic field $\boldsymbol{B}$ in the direction perpendicular to the flow, with a maximum field strength of $0.3\,\mathrm{T}$. An electromagnetic pump maintains a constant volumetric flow rate at the channel inlet, and the free-surface height is measured along the channel using a laser-camera system. The liquid metal is Galinstan (GaInSn) at room temperature,\cite{MorleyRSI08} and the channel liner is modeled as copper. Under the applied field, the $\boldsymbol{j}\times\boldsymbol{B}$ Lorentz force retards the flow, causing liquid metal pileup that manifests as a monotonically decreasing free-surface height along the channel. Additional experimental parameters are reported in Ref.~\onlinecite{SunNF23}.

\begin{figure}[htbp]
  \centering
  \includegraphics[width=\linewidth]{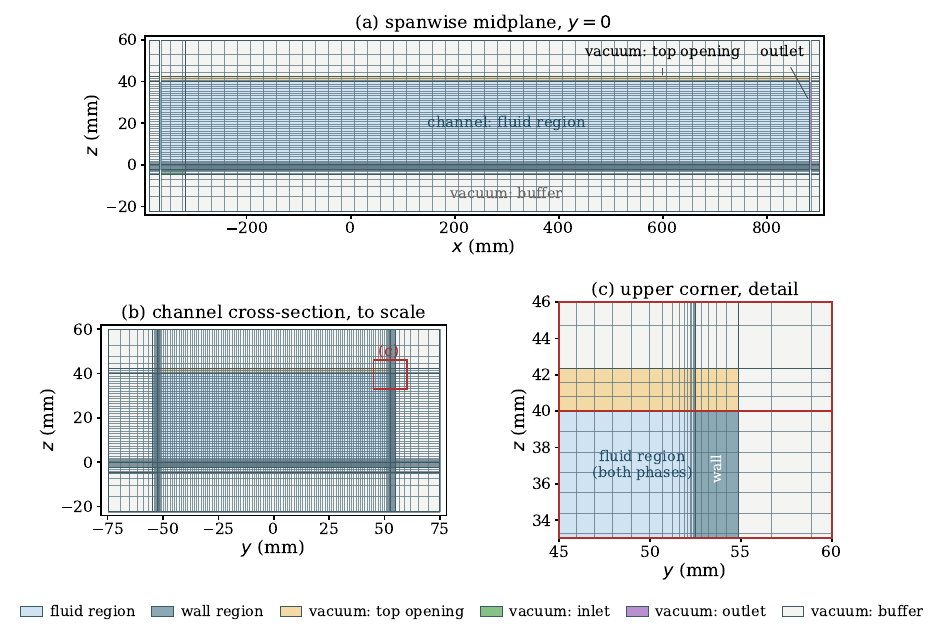}
  \caption{\label{fig:regions}Region decomposition and mesh of the LMX-U case, drawn from the computational mesh. (a)~Spanwise midplane, $y=0$, vertical scale exaggerated four-fold, every twenty-fifth streamwise cell line drawn; (b)~channel cross-section, to scale; (c)~detail of the boxed upper corner at full cell resolution, where the fluid-region cells run to the top of the channel and the red line marks the vacuum region above it. The free surface is not drawn.}
\end{figure}

The simulation domain reproduces the LMX-U geometry and is initialized with the following parameters: volumetric flow rate of $\langle Q \rangle = 2.72\times 10^{-4}\,\mathrm{m^3\,s^{-1}}$ ($16.3\,\mathrm{L\,min^{-1}}$), Hartmann number $Ha = 4.1\times 10^2$ at $0.2\,\mathrm{T}$ and $6.1\times 10^2$ at $0.3\,\mathrm{T}$, and magnetic Reynolds number $R_m = 3.5\times10^{-2}$. The external magnetic field is applied as a uniform transverse field over the magnet region, with strengths of $0.2\,\mathrm{T}$ and $0.3\,\mathrm{T}$ corresponding to the two experimental cases. Material properties of Galinstan at room temperature are taken from Ref.~\onlinecite{MorleyRSI08}, and the copper liner conductivity is assumed to be $\sigma_w = 59.98\times10^6\,\mathrm{S\,m^{-1}}$. The resulting mesh and region decomposition are shown in Fig.~\ref{fig:regions}. The channel is meshed as a single fluid region whose cells run continuously from the floor to the top of the channel, so that the liquid metal and the gas above it share one mesh and the free surface lies within it; the copper liner is a separate solid region, and four vacuum regions surround the assembly, one of them occupying the space above the open top of the channel. Neighbouring regions meet at mapped interfaces rather than at a shared mesh. The hydrodynamic and interface conditions are as follows. Liquid metal enters through the inlet region at the fixed volumetric flow rate quoted above, with the phase fraction set to unity there; the channel floor and the side walls are no-slip and impermeable; and the outlet and the open top of the channel carry a combined pressure--velocity condition that admits inflow or outflow according to the local pressure, the reference pressure being fixed at the outlet. The phase fraction is zero-gradient on the walls, at the outlet and at the open top, so no contact-angle condition is imposed and the interface meets the walls at $90^\circ$. The surface tension is $\sigma=0.375\,\mathrm{N\,m^{-1}}$.

The mesh resolves the full $105\,\mathrm{mm}$ span, with around $120$ cells across; no symmetry plane or symmetry patch is used anywhere in the case. It carries approximately ten million cells in total, of which $60\%$ are in the fluid region.
 
\begin{figure}[htbp]
  \centering
  \includegraphics[width=\linewidth]{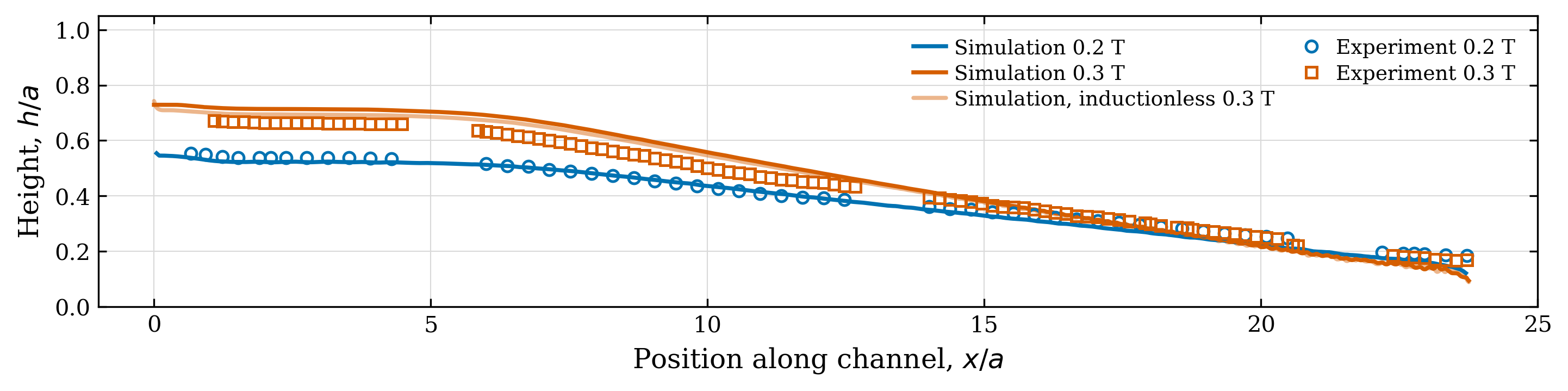}
  \caption{Time-averaged free-surface height profiles along the LMX-U channel at $0.2\,\mathrm{T}$ and $0.3\,\mathrm{T}$, at a fixed volumetric flow rate $\langle Q \rangle = 2.72\times10^{-4}\, \mathrm{m^3\,s^{-1}}$. Height $h$ and position $x$ are normalized by the channel half-width $a$, with the vertical scale exaggerated five-fold. Solid lines are the computed profiles with the induced field resolved; the translucent line is the corresponding computation at $0.3\,\mathrm{T}$ with induction neglected, obtained with the original inductionless \texttt{FreeMHD} solver\cite{WynnePoP25}. Markers are the measurements of Ref.~\onlinecite{SunNF23}. The computed profiles are averaged over the quasi-steady window, the measured thickness being itself a moving average through the individual frames.\cite{SunNF23}}
  \label{LMX U}
\end{figure}

Fig.~\ref{LMX U} compares the simulated steady-state free-surface height profiles with experimental measurements at external field strengths of $0.2\,\mathrm{T}$ and $0.3\,\mathrm{T}$. Solid lines represent simulation results, and markers denote experimental data. In both cases the solver correctly captures the qualitative trend: the free surface slopes downward along the channel as MHD drag decelerates the flow and drives liquid metal accumulation near the inlet. The quantitative agreement is good, particularly in the downstream portion of the channel where the height gradient is most pronounced. At the low magnetic Reynolds number of this experiment, the height profile computed with the induced field resolved nearly coincides with the corresponding inductionless computation, as it must: the induced field is dynamically negligible here, and the coincidence constitutes a consistency check that the resolved formulation correctly reduces to the inductionless limit, rather than a redundancy of the extension. The validation at this $R_m$ therefore assesses the overall free-surface MHD solution; the correctness of the resolved induced-field dynamics is established by the verification cases of Secs.~\ref{sec:analytical_verification} and~\ref{sec:transient_verification}.

\begin{figure}[htbp]
  \centering
  \includegraphics[width=\linewidth]{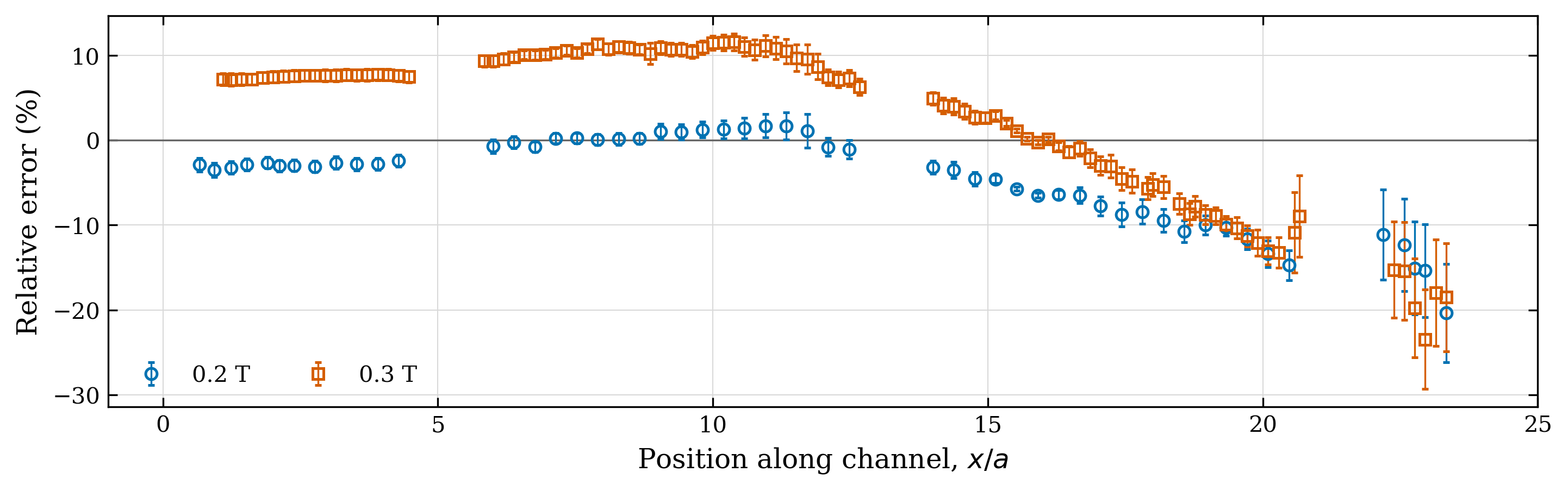}
  \caption{Signed relative error of the computed free-surface height, $(h_{\mathrm{sim}}-h_{\mathrm{exp}})/h_{\mathrm{exp}}$, at the experimental sampling positions of Fig.~\ref{LMX U}, for $0.2\,\mathrm{T}$ (circles) and $0.3\,\mathrm{T}$ (squares). Error bars are the reported measurement uncertainty propagated into the relative error. The gaps reflect the coverage of the laser--camera diagnostic.}
  \label{LMX err}
\end{figure}

The difference between the simulated and measured heights is quantified in Fig.~\ref{LMX err}, which shows the signed relative error $(h_{\mathrm{sim}}-h_{\mathrm{exp}})/h_{\mathrm{exp}}$ at the experimental sampling positions. The error is not a single systematic offset. At $0.2\,\mathrm{T}$ the computed height lies about $3\%$ below the measurement over the first few half-widths, agrees with it to within $1.5\%$ between $x/a\approx6$ and $x/a\approx12$, and falls progressively below it thereafter; at $0.3\,\mathrm{T}$ it lies $7$--$11\%$ above the measurement upstream of $x/a\approx16$; both fall below the measurement in the last few half-widths before the outlet, where the film is thinnest. The root-mean-square error over the channel is $7.1\%$ at $0.2\,\mathrm{T}$ and $9.4\%$ at $0.3\,\mathrm{T}$. The reported measurement uncertainty is not uniform along the channel: it averages $0.7\%$ of the local film height over the upstream third, $0.9\%$ over the middle third, and $2.6\%$ over the last third where the film is thinnest, reaching $6.9\%$ at individual stations.

Several sources may contribute. The height responds more strongly to the applied field in the computation than in the experiment, the signature of an over-estimated wall conductance, which progressive oxidation of the copper liner\cite{WynnePoP25} would produce. Superposed on that is a field-independent offset, to which the imposed inlet flow rate is the natural candidate. In the experiment, the flow rate was controlled not by a direct measurement, but by maintaining the pump rate. On the numerical modeling side, the applied field is prescribed as a purely transverse component, so the perpendicular components a physical magnet produces where the field varies, and the currents they would drive, are absent. For more information on the experimental configuration and its uncertainties, please refer to Ref.~\onlinecite{SunNF23}.

Two checks bound the numerical contribution to that difference. Mesh independence is assessed at $0.2\,\mathrm{T}$ by repeating the calculation on an independently constructed mesh. One is an unstructured, hex-dominant mesh generated by \texttt{snappyHexMesh} ($9.05\times10^{6}$ cells) with local refinement and near-wall layers; the other is block-structured, built by \texttt{blockMesh} ($10.7\times10^{6}$ cells) and graded toward the walls. Their height profiles, each averaged over its quasi-steady window, differ by $2.1\%$ over the channel. Time-step sensitivity is assessed on the \texttt{snappyHexMesh} mesh by halving both the Courant limit and the maximum time step, which changes the window-averaged profile by $0.34\%$ over the channel and by $0.06\%$ over the upstream third. The two meshes differ in generator, topology and cell count, so their agreement bounds the discretization error more broadly than a nested refinement of a single mesh would. Both figures are far smaller than the difference from the measurements, which is therefore not a discretization artefact.

\begin{figure}
\includegraphics[width=\linewidth]{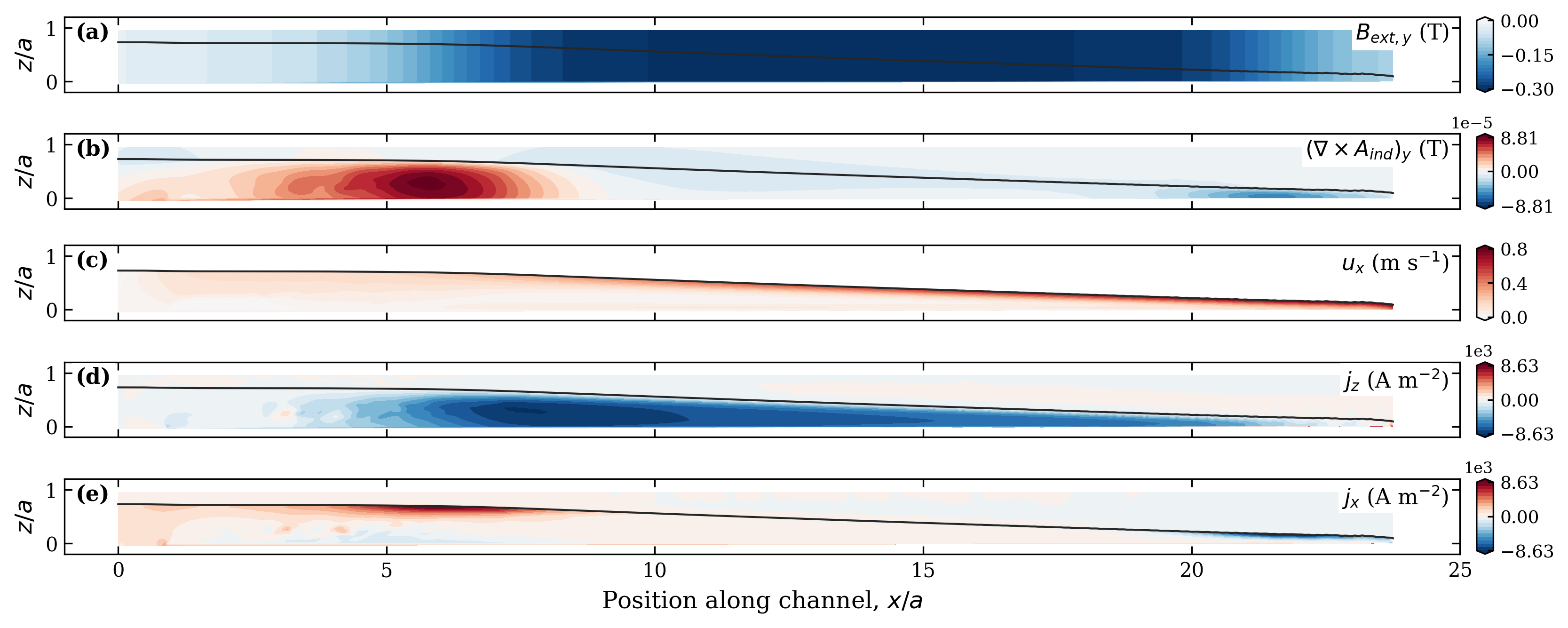}
\caption{\label{LMX B}Steady-state field, flow, and current structure in the LMX-U simulation at $B_{\mathrm{ext}}=0.3\,\mathrm{T}$, on the vertical midplane ($x$-$z$ plane, $y$ into the page). (a)~Applied field component $B_{\mathrm{ext},y}$; (b)~induced field component $(\nabla\times\boldsymbol{A}_{\mathrm{ind}})_y$; (c)~streamwise velocity $u_x$, drawn in the liquid metal only, the entrained gas being blanked; (d)~vertical current density $j_z$; (e)~streamwise current density $j_x$. Panels (d) and (e) share a color scale. The solid black line marks the free surface.}
\end{figure}

The steady-state electromagnetic and flow structure underlying the height profiles of Fig.~\ref{LMX U} is examined in Fig.~\ref{LMX B} for the $0.3\,\mathrm{T}$ case. The external field applied by the electromagnet [Fig.~\ref{LMX B}(a)] is uniform over the magnet span but exhibits strong streamwise gradients in the fringe regions near the inlet and outlet ends of the magnet. Within the magnetized region, the motional electromotive force drives a vertical current $j_z$ [Fig.~\ref{LMX B}(d)] that occupies the bulk of the conducting fluid, and the associated $\boldsymbol{j}\times\boldsymbol{B}$ force decelerates the flow in the negative $x$-direction. Because liquid metal is continuously injected at a fixed flow rate, the steady state that results carries the majority of the volumetric flow in a high-speed layer beneath the free surface [Fig.~\ref{LMX B}(c)], while the bulk beneath it is strongly retarded.

\begin{figure}
\includegraphics[width=\linewidth]{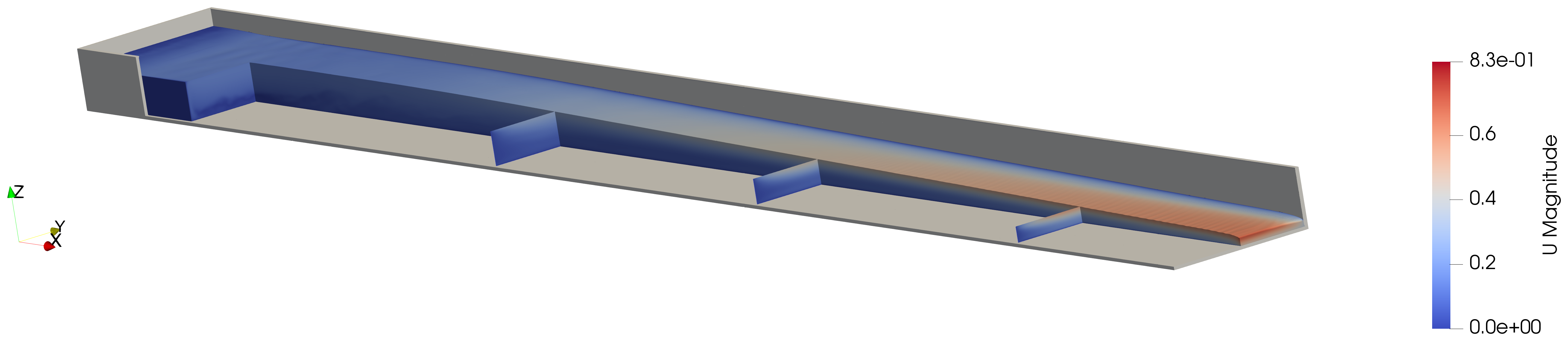}
\caption{\label{LMX 3D}Three-dimensional view of the LMX-U simulation at $B_{\mathrm{ext}}=0.2\,\mathrm{T}$, with the near side wall cut away and the channel walls drawn in grey. The free surface and four cross-stream planes are colored by velocity magnitude, the flow running from the inlet basin at the left to the outlet at the right; the gas above the free surface is masked, as in Fig.~\ref{LMX B}. The full $105\,\mathrm{mm}$ width is resolved, with no symmetry plane.}
\end{figure}

\begin{figure}
\includegraphics[width=\linewidth]{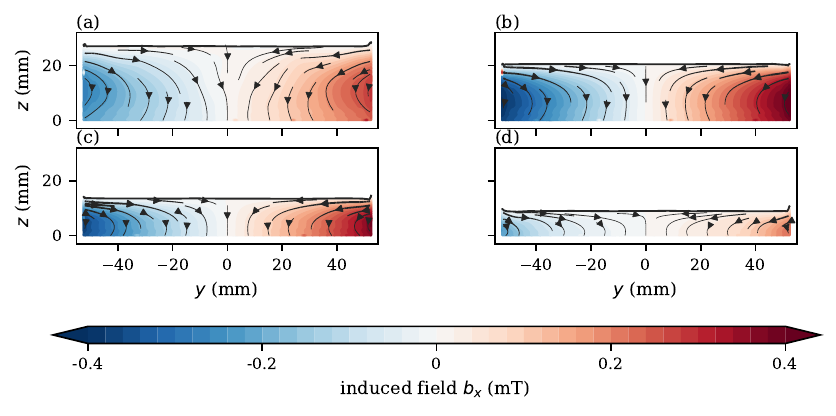}
\caption{\label{LMX xsec}Induced field and current density on $y$--$z$ cross-sections of the LMX-U channel at $B_{\mathrm{ext}}=0.2\,\mathrm{T}$ and $t=20.9\,\mathrm{s}$. Panels (a)--(d) are at $x=-0.07$, $+0.27$, $+0.60$ and $+0.81\,\mathrm{m}$, the first and last on the fringes of the applied field and the middle two in its uniform core. Color is the streamwise induced field $b_x=(\nabla\times\boldsymbol{A}_{\mathrm{ind}})_x$. Streamlines are the in-plane current density, and the dark line is the $\alpha=0.5$ free surface.}
\end{figure}

The self-consistently resolved induced field [Fig.~\ref{LMX B}(b)], a quantity that is unavailable in the inductionless formulation, reveals additional structure in this nonlinearly coupled system. The induced field is strongly localized in the fringe regions of the external field, with opposite polarity at the inlet and outlet ends, following the sign reversal of the streamwise field gradient. This localization reflects the closure of the induced current system: the non-uniform motional electromotive force in the fringe regions drives current loops in the $x$-$z$ plane, whose streamwise legs appear as the $j_x$ distribution of Fig.~\ref{LMX B}(e) and whose circulation generates the $y$-directed induced field. Although $j_x$ and $j_z$ are plotted on the same color scale, their spatial characters differ markedly: $j_z$ is a bulk current, whereas $j_x$ is confined largely to the free surface and the boundary layers. The induced field magnitude is more than three orders of magnitude smaller than the applied field, consistent with the low magnetic Reynolds number of this configuration. Figure~\ref{LMX 3D} shows the same calculation in three dimensions, the free surface and four cross-stream planes colored by velocity magnitude, the full channel width being resolved with no symmetry plane. For the $0.2\,\mathrm{T}$ case, Fig.~\ref{LMX xsec} resolves that localization across the channel rather than on the midplane alone, and shows the current closing in loops within the plane, with the return legs running beneath the free surface, so that the circuit is contained in no single streamwise plane.

The numerical diffusion of the captured interface is quantified directly for the $0.2\,\mathrm{T}$ case. On the mesh used, the vertical cell size near the interface is $1.13\,\mathrm{mm}$ and the phase fraction passes from $\alpha=0.95$ to $\alpha=0.05$ over about three cells, i.e.\ over $3.2\text{--}3.4\,\mathrm{mm}$, at every streamwise station. The absolute thickness of the transition layer is therefore constant along the channel, as expected for algebraic interface capturing with compression; the apparent downstream increase of interfacial diffusion is a consequence of the thinning film, whose height falls from $h/a\approx0.53$ near the inlet to $h/a\approx0.17$ near the outlet, so that the same band occupies $12\%$ of the local film height at $x/a=1$ and $36\%$ at $x/a=22$.

These observations constitute a preliminary finding rather than a systematic study. Although dynamically negligible at the low $R_m$ of LMX-U, both features, the induced field localized in the external field gradient regions and the return current concentrated at the free surface, can be expected to strengthen as $R_m$ increases toward the finite-$R_m$ conditions anticipated in reactor environments, where large transient electromagnetic events and high-throughput liquid metal concepts push the flow out of the inductionless regime. Their reactor relevance is twofold. First, sharp effective field gradients can arise in the frame of the flowing liquid during transient electromagnetic events: edge-localized modes, vertical displacement events, and disruptions impose rapid spatio-temporal field variations on the liquid metal, for which the fringe regions serve as a stationary, experimentally accessible analogue of a strongly non-uniform field seen by the moving conductor. Second, currents concentrated at the free surface couple directly to the stability of the liquid metal interface, which in turn governs the surface heat flux handling of a liquid metal PFC. Both features therefore motivate the systematic study of free-surface liquid metal flows with resolved induced fields identified as future work in Sec.~\ref{conclusion}.

\subsection{Free-surface deformation of a liquid metal jet}\label{sec:jet}

The verification cases of Sec.~\ref{sec:analytical_verification} and Sec.~\ref{sec:transient_verification} are closed geometries, and the LMX-U comparison of Sec.~\ref{sec:experimental_validation} is a channel flow whose free surface remains a single-valued height profile. Neither exercises the free-surface treatment in a configuration where the interface deforms substantially in both transverse directions. This section adds such a case: the mercury jet of Samulyak~\textit{et al.}\cite{SamulyakJCP07}, Sec.~4, which enters a strongly non-uniform transverse magnetic field and is deformed by the resulting Lorentz force.

A cylindrical jet of mercury of radius $R_0=4\,\mathrm{mm}$ travels at $u=60\,\mathrm{m\,s^{-1}}$ into a static transverse field whose magnitude varies along the direction of travel as $(B_y/B_{\max})^2=\tfrac{1}{2}\left[1-\tanh\left((z-z_0)/L_m\right)\right]$, with $z_0=1.5\,\mathrm{cm}$ and $L_m=0.62\,\mathrm{cm}$. The currents induced as the jet crosses the field gradient interact with the field to compress the jet along the field-perpendicular transverse axis and, by incompressibility, to expand it along the field direction. We use mercury properties $\sigma=1.04\times10^{6}\,\mathrm{S\,m^{-1}}$, $\rho=13\,534\,\mathrm{kg\,m^{-3}}$, $\mu=1.526\times10^{-3}\,\mathrm{Pa\,s}$ and surface tension $\sigma_s=0.485\,\mathrm{N\,m^{-1}}$, giving a Weber number $We=\rho u^{2}R_0/\sigma_s\approx4\times10^{5}$, so surface tension is negligible here. The magnetic Reynolds number formed with the jet radius is $R_m=\mu_0\sigma uR_0=0.31$. The physical permeability of free space is used.

The calculation is performed in the laboratory frame, with the jet advecting through a static field; the reference held the jet fixed and translated the field instead. The fluid region spans $\pm7.5\,\mathrm{mm}$ across the compressed axis and $\pm15\,\mathrm{mm}$ across the expanded one, discretized with $0.5\,\mathrm{mm}$ cubic cells ($648\,000$ cells), and is embedded in a graded vacuum region extending to $\pm40\,\mathrm{mm}$ with $\boldsymbol{A}=0$ on its outer boundary ($1.44\times10^{6}$ cells in total). The jet is $6\,\mathrm{cm}$ long with both ends interior to the domain; stations within $2.5R_0$ of either end are excluded from the comparison, following the reference's own restriction. The deformation is measured from $\alpha$-weighted second moments of each cross-section, with $R_0$ taken from the discrete initial field.

\begin{figure}[htbp]
  \centering
  \includegraphics[width=\linewidth]{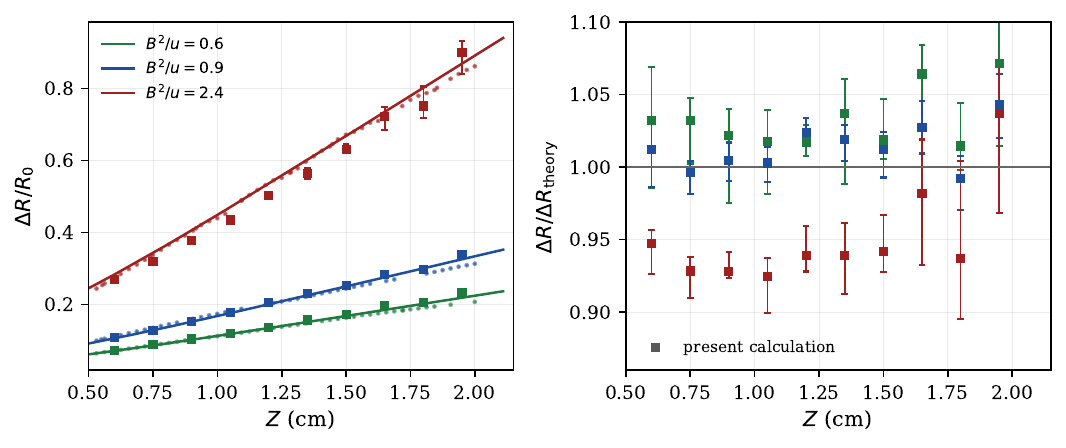}
  \caption{\label{jet deformation}Deformation of a mercury jet entering a non-uniform transverse magnetic field, compared with Fig.~10 of Ref.~\onlinecite{SamulyakJCP07}. (Left) Deformation along the field direction against distance travelled past the field center, at three field strengths: solid lines are the reference's theory, small dots its simulation data, and squares the present calculation. (Right) The same data as a ratio to that theory. Bars are the full spread over the nine sampled times in the quasi-steady window.}
\end{figure}

\begin{figure}[htbp]
  \centering
  \includegraphics[width=0.62\linewidth]{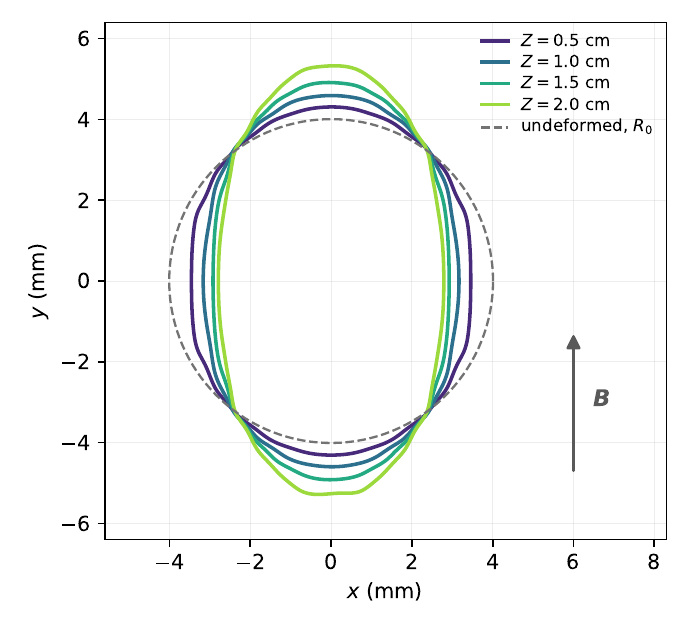}
  \caption{\label{jet xsec}The $\alpha=0.5$ interface of the mercury jet at $B_{\max}^2/u=0.9\,\mathrm{T^2\,s\,m^{-1}}$, on cross-sections at four distances $Z$ past the field centre, with the undeformed circle of radius $R_0$ dashed. The phase fraction is averaged over the same nine sampled times of the quasi-steady window as the deformations of Table~\ref{jet table}. The applied field lies along $y$. The deformations of Fig.~\ref{jet deformation} and Table~\ref{jet table} are obtained from $\alpha$-weighted second moments rather than from this contour.}
\end{figure}

Figure~\ref{jet deformation} compares the computed deformation with the theory and the simulation data of Ref.~\onlinecite{SamulyakJCP07}, Fig.~10, at three field strengths parameterized by $B_{\max}^2/u$. Table~\ref{jet table} collects the comparison at $Z=1\,\mathrm{cm}$ past the field center. At $B_{\max}^2/u=0.6$ and $0.9\,\mathrm{T^2\,s\,m^{-1}}$ the computed deformation is within $1.1\%$ and $0.4\%$ of the reference theory, and the computed points lie on the reference's own simulation data as well as on its analytical curve. The reference requires the deformation at a fixed station to be time-invariant for the comparison to be meaningful; over the sampling window $t=0.85$--$1.25\,\mathrm{ms}$ ours varies by $2$--$4\%$ peak to peak, so that criterion is satisfied. Figure~\ref{jet xsec} shows the interface at four stations, the initially circular jet flattening along the field direction as it advances.

\begin{table}[htbp]
\caption{\label{jet table}Jet deformation $\Delta R/R_0$ along the field direction at $Z=1\,\mathrm{cm}$ past the field center, against the theory of Ref.~\onlinecite{SamulyakJCP07}, Fig.~10. Values are means over the nine sampled times in the quasi-steady window, with standard errors of that mean. (Sec.~\ref{sec:jet})}
\begin{ruledtabular}
\begin{tabular}{lccc}
$B_{\max}^2/u$ (T$^2$\,s\,m$^{-1}$) & $B_{\max}$ (T) & $\Delta R/R_0$ & Ratio to Ref.~\onlinecite{SamulyakJCP07} \\
\hline
$0.6$ & $6.00$  & $0.1146$ & $1.011\pm0.004$ \\
$0.9$ & $7.35$  & $0.1679$ & $0.996\pm0.002$ \\
$2.4$ & $12.00$ & $0.4144$ & $0.923\pm0.004$ \\
\end{tabular}
\end{ruledtabular}
\end{table}

The precision of that agreement is set by the representation of the interface. Ref.~\onlinecite{SamulyakJCP07} tracks the interface explicitly, whereas our solver captures it algebraically on the Eulerian mesh, so that the phase fraction passes from $\alpha=0.95$ to $\alpha=0.05$ over three cells, $1.5\,\mathrm{mm}$, against a jet radius of $4\,\mathrm{mm}$. The deformation is area-preserving to within $3\%$: the product of the semi-axes of the $\alpha=0.5$ contour of Fig.~\ref{jet xsec} lies between $0.97$ and $1.00$ times $R_0^2$ over the comparison window. The $\alpha$-weighted second moments from which the tabulated deformations are obtained differ from that contour by up to four percentage points in either semi-axis, and in opposite senses, which is the precision the values quoted above carry.

\section{Discussion and Conclusion}\label{conclusion}

This work has extended \texttt{FreeMHD} beyond the inductionless approximation. The resulting solver resolves the evolution of the induced magnetic field in three-dimensional, transient, free-surface liquid metal flows, filling a gap that had persisted despite a growing need in fusion reactor design.

The analytical verification against the Shercliff and Hunt duct-flow solutions demonstrates that the electromagnetic and momentum-transport implementations are correct across a wide range of Hartmann numbers. The two cases are complementary: the Shercliff case, with insulating Hartmann walls, confirms accurate resolution of the induced-field profile and the Hartmann and side-wall boundary layers in the absence of wall currents; the Hunt case, with conducting Hartmann walls, additionally tests the solver's ability to reproduce the wall-current-driven side-wall jets and the more complex induced-field topology that results. Agreement with the analytical solutions of Ref.~\onlinecite{MullerBook} is good in both cases across all three Hartmann numbers examined. The subsequent experimental validation against LMX-U free-surface height measurements provides confidence that the solver performs correctly in a realistic, multi-region, free-surface configuration with the induced field resolved. Because this experiment operates at $R_m\sim O(10^{-2})$, where the resolved and inductionless predictions nearly coincide, it validates the overall free-surface MHD solver rather than finite-$R_m$ induced-field effects, for which no free-surface experimental data are presently available. Beyond the validation itself, the resolved solution reveals induced magnetic fields localized in the external field gradient regions and return currents concentrated at the free surface, features that are invisible to inductionless solvers. These are preliminary observations at low $R_m$, but they merit dedicated study of free-surface liquid metal flows with induced fields.

By resolving the induced field directly, the solver provides a quantitative basis for assessing when the inductionless approximation holds and when it does not. The verification and validation reported here span $R_m$ from $10^{-3}$ to $4.6$, the range for which analytical and experimental references are available. Within that range, the transient-field case of Sec.~\ref{sec:transient_verification} demonstrates more than the correctness of the induced-field solution: the flow there is generated entirely by the time variation of the applied field, through an electromotive force that the inductionless formulation does not contain at any $R_m$, so the solver reproduces the class of physics for which resolving the induced field is indispensable rather than merely advantageous. That case also carries the calculation through $R_m\sim\mathcal{O}(1)$ and beyond, where the coupling ceases to be one-way. A free-surface counterpart is provided at lower $R_m$ by the deforming jet of Sec.~\ref{sec:jet}, whose deformation reproduces the reference theory. What remains outstanding is the combination of the two: a free-surface flow at $R_m\sim\mathcal{O}(1)$, for which no analytical solution or experimental dataset presently exists. The fringe-localized induced fields and free-surface currents identified in Sec.~\ref{sec:experimental_validation} provide natural starting points for such studies, since both are expected to strengthen with $R_m$ and both bear directly on the field gradient and free-surface stability questions that arise in reactor environments.

Two extensions are deferred to future work. Heat transport, currently under testing, will enable fully multiphysics simulations under reactor-relevant thermal and electromagnetic loads. A more general treatment of the vacuum exterior boundary, here closed by a surrounding region with $\boldsymbol{A}_{\mathrm{ind}}=\boldsymbol{0}$ on its outer surface, will be reported separately.

\begin{acknowledgments}
This research was supported by the U.S. Department of Energy under Award DE-SC0024626, and by R\&D Program of ``Optimal Basic Design of DEMO Fusion Reactor, CN2602-2'' through the Korea Institute of Fusion Energy (KFE) funded by the Government funds. The authors gratefully acknowledge the Research Institute of Energy and Resources and the Institute of Engineering Research at Seoul National University.
\end{acknowledgments}

\section*{Data Availability}
The source code is openly available in the \texttt{FreeMHD} GitHub repository,\cite{FreeMHD} and the simulation input files are available in the Zenodo repository.\cite{Zenodo}

\bibliography{reference}

\begin{thebibliography}{43}%
\makeatletter
\providecommand \@ifxundefined [1]{%
 \@ifx{#1\undefined}
}%
\providecommand \@ifnum [1]{%
 \ifnum #1\expandafter \@firstoftwo
 \else \expandafter \@secondoftwo
 \fi
}%
\providecommand \@ifx [1]{%
 \ifx #1\expandafter \@firstoftwo
 \else \expandafter \@secondoftwo
 \fi
}%
\providecommand \natexlab [1]{#1}%
\providecommand \enquote  [1]{``#1''}%
\providecommand \bibnamefont  [1]{#1}%
\providecommand \bibfnamefont [1]{#1}%
\providecommand \citenamefont [1]{#1}%
\providecommand \href@noop [0]{\@secondoftwo}%
\providecommand \href [0]{\begingroup \@sanitize@url \@href}%
\providecommand \@href[1]{\@@startlink{#1}\@@href}%
\providecommand \@@href[1]{\endgroup#1\@@endlink}%
\providecommand \@sanitize@url [0]{\catcode `\\12\catcode `\$12\catcode
  `\&12\catcode `\#12\catcode `\^12\catcode `\_12\catcode `\%12\relax}%
\providecommand \@@startlink[1]{}%
\providecommand \@@endlink[0]{}%
\providecommand \url  [0]{\begingroup\@sanitize@url \@url }%
\providecommand \@url [1]{\endgroup\@href {#1}{\urlprefix }}%
\providecommand \urlprefix  [0]{URL }%
\providecommand \Eprint [0]{\href }%
\providecommand \doibase [0]{http://dx.doi.org/}%
\providecommand \selectlanguage [0]{\@gobble}%
\providecommand \bibinfo  [0]{\@secondoftwo}%
\providecommand \bibfield  [0]{\@secondoftwo}%
\providecommand \translation [1]{[#1]}%
\providecommand \BibitemOpen [0]{}%
\providecommand \bibitemStop [0]{}%
\providecommand \bibitemNoStop [0]{.\EOS\space}%
\providecommand \EOS [0]{\spacefactor3000\relax}%
\providecommand \BibitemShut  [1]{\csname bibitem#1\endcsname}%
\let\auto@bib@innerbib\@empty
\bibitem [{\citenamefont {Haiqi}\ \emph {et~al.}(2008)\citenamefont {Haiqi},
  \citenamefont {Baofeng}, \citenamefont {Huiqin},\ and\ \citenamefont
  {Jianchao}}]{HaiqiJMPT08}%
  \BibitemOpen
  \bibfield  {author} {\bibinfo {author} {\bibfnamefont {Y.}~\bibnamefont
  {Haiqi}}, \bibinfo {author} {\bibfnamefont {W.}~\bibnamefont {Baofeng}},
  \bibinfo {author} {\bibfnamefont {L.}~\bibnamefont {Huiqin}}, \ and\ \bibinfo
  {author} {\bibfnamefont {L.}~\bibnamefont {Jianchao}},\ }\href {\doibase
  10.1016/j.jmatprotec.2007.08.054} {\bibfield  {journal} {\bibinfo  {journal}
  {Journal of Materials Processing Technology}\ }\textbf {\bibinfo {volume}
  {202}},\ \bibinfo {pages} {179} (\bibinfo {year} {2008})}\BibitemShut
  {NoStop}%
\bibitem [{\citenamefont {Zhang}\ \emph {et~al.}(2021)\citenamefont {Zhang},
  \citenamefont {Liu}, \citenamefont {Fan}, \citenamefont {Zhang},
  \citenamefont {Zhou}, \citenamefont {Kalantar-Zadeh},\ and\ \citenamefont
  {Guo}}]{ZhangEES21}%
  \BibitemOpen
  \bibfield  {author} {\bibinfo {author} {\bibfnamefont {S.}~\bibnamefont
  {Zhang}}, \bibinfo {author} {\bibfnamefont {Y.}~\bibnamefont {Liu}}, \bibinfo
  {author} {\bibfnamefont {Q.}~\bibnamefont {Fan}}, \bibinfo {author}
  {\bibfnamefont {C.}~\bibnamefont {Zhang}}, \bibinfo {author} {\bibfnamefont
  {T.}~\bibnamefont {Zhou}}, \bibinfo {author} {\bibfnamefont {K.}~\bibnamefont
  {Kalantar-Zadeh}}, \ and\ \bibinfo {author} {\bibfnamefont {Z.}~\bibnamefont
  {Guo}},\ }\href {\doibase 10.1039/D1EE00531F} {\bibfield  {journal} {\bibinfo
   {journal} {Energy \& Environmental Science}\ }\textbf {\bibinfo {volume}
  {14}},\ \bibinfo {pages} {4177} (\bibinfo {year} {2021})}\BibitemShut
  {NoStop}%
\bibitem [{\citenamefont {Xiang}\ \emph {et~al.}(2024)\citenamefont {Xiang},
  \citenamefont {Yang}, \citenamefont {Wang},\ and\ \citenamefont
  {Wu}}]{XiangTSEP24}%
  \BibitemOpen
  \bibfield  {author} {\bibinfo {author} {\bibfnamefont {L.}~\bibnamefont
  {Xiang}}, \bibinfo {author} {\bibfnamefont {S.}~\bibnamefont {Yang}},
  \bibinfo {author} {\bibfnamefont {Q.}~\bibnamefont {Wang}}, \ and\ \bibinfo
  {author} {\bibfnamefont {J.}~\bibnamefont {Wu}},\ }\href {\doibase
  10.1016/j.tsep.2024.102560} {\bibfield  {journal} {\bibinfo  {journal}
  {Thermal Science and Engineering Progress}\ }\textbf {\bibinfo {volume}
  {50}},\ \bibinfo {pages} {102560} (\bibinfo {year} {2024})}\BibitemShut
  {NoStop}%
\bibitem [{\citenamefont {Cortazar}\ and\ \citenamefont
  {Lazarus}(2025)}]{CortazarAMT25}%
  \BibitemOpen
  \bibfield  {author} {\bibinfo {author} {\bibfnamefont {J.}~\bibnamefont
  {Cortazar}}\ and\ \bibinfo {author} {\bibfnamefont {N.}~\bibnamefont
  {Lazarus}},\ }\href {\doibase 10.1002/admt.202401372} {\bibfield  {journal}
  {\bibinfo  {journal} {Advanced Materials Technologies}\ }\textbf {\bibinfo
  {volume} {10}},\ \bibinfo {pages} {2401372} (\bibinfo {year}
  {2025})}\BibitemShut {NoStop}%
\bibitem [{\citenamefont {Yang}\ \emph {et~al.}(2007)\citenamefont {Yang},
  \citenamefont {Ma}, \citenamefont {Bliss},\ and\ \citenamefont
  {Bryant}}]{YangIJHFF07}%
  \BibitemOpen
  \bibfield  {author} {\bibinfo {author} {\bibfnamefont {M.}~\bibnamefont
  {Yang}}, \bibinfo {author} {\bibfnamefont {N.}~\bibnamefont {Ma}}, \bibinfo
  {author} {\bibfnamefont {D.}~\bibnamefont {Bliss}}, \ and\ \bibinfo {author}
  {\bibfnamefont {G.}~\bibnamefont {Bryant}},\ }\href {\doibase
  10.1016/j.ijheatfluidflow.2006.08.001} {\bibfield  {journal} {\bibinfo
  {journal} {International Journal of Heat and Fluid Flow}\ }\textbf {\bibinfo
  {volume} {28}},\ \bibinfo {pages} {768} (\bibinfo {year} {2007})}\BibitemShut
  {NoStop}%
\bibitem [{\citenamefont {Miyazawa}\ \emph {et~al.}(2017)\citenamefont
  {Miyazawa}, \citenamefont {Goto}, \citenamefont {Murase}, \citenamefont
  {Ohgo}, \citenamefont {Yanagi}, \citenamefont {Tanaka}, \citenamefont
  {Tamura}, \citenamefont {Tanaka}, \citenamefont {Masuzaki}, \citenamefont
  {Sakamoto}, \citenamefont {Yagi}, \citenamefont {Sagara},\ and\ \citenamefont
  {the FFHR Design~Group}}]{MiyazawaFED17}%
  \BibitemOpen
  \bibfield  {author} {\bibinfo {author} {\bibfnamefont {J.}~\bibnamefont
  {Miyazawa}}, \bibinfo {author} {\bibfnamefont {T.}~\bibnamefont {Goto}},
  \bibinfo {author} {\bibfnamefont {T.}~\bibnamefont {Murase}}, \bibinfo
  {author} {\bibfnamefont {T.}~\bibnamefont {Ohgo}}, \bibinfo {author}
  {\bibfnamefont {N.}~\bibnamefont {Yanagi}}, \bibinfo {author} {\bibfnamefont
  {H.}~\bibnamefont {Tanaka}}, \bibinfo {author} {\bibfnamefont
  {H.}~\bibnamefont {Tamura}}, \bibinfo {author} {\bibfnamefont
  {T.}~\bibnamefont {Tanaka}}, \bibinfo {author} {\bibfnamefont
  {S.}~\bibnamefont {Masuzaki}}, \bibinfo {author} {\bibfnamefont
  {R.}~\bibnamefont {Sakamoto}}, \bibinfo {author} {\bibfnamefont
  {J.}~\bibnamefont {Yagi}}, \bibinfo {author} {\bibfnamefont {A.}~\bibnamefont
  {Sagara}}, \ and\ \bibinfo {author} {\bibnamefont {the FFHR Design~Group}},\
  }\href {\doibase 10.1016/j.fusengdes.2017.07.003} {\bibfield  {journal}
  {\bibinfo  {journal} {Fusion Engineering and Design}\ }\textbf {\bibinfo
  {volume} {125}},\ \bibinfo {pages} {227} (\bibinfo {year}
  {2017})}\BibitemShut {NoStop}%
\bibitem [{\citenamefont {Alegre}\ \emph {et~al.}(2020)\citenamefont {Alegre},
  \citenamefont {Oyarzabal}, \citenamefont {Tafalla}, \citenamefont {Liniers},
  \citenamefont {Soleto},\ and\ \citenamefont {Tabar{\'e}s}}]{AlegreJFE20}%
  \BibitemOpen
  \bibfield  {author} {\bibinfo {author} {\bibfnamefont {D.}~\bibnamefont
  {Alegre}}, \bibinfo {author} {\bibfnamefont {E.}~\bibnamefont {Oyarzabal}},
  \bibinfo {author} {\bibfnamefont {D.}~\bibnamefont {Tafalla}}, \bibinfo
  {author} {\bibfnamefont {M.}~\bibnamefont {Liniers}}, \bibinfo {author}
  {\bibfnamefont {A.}~\bibnamefont {Soleto}}, \ and\ \bibinfo {author}
  {\bibfnamefont {F.}~\bibnamefont {Tabar{\'e}s}},\ }\href {\doibase
  10.1007/s10894-020-00254-5} {\bibfield  {journal} {\bibinfo  {journal}
  {Journal of Fusion Energy}\ }\textbf {\bibinfo {volume} {39}},\ \bibinfo
  {pages} {411} (\bibinfo {year} {2020})}\BibitemShut {NoStop}%
\bibitem [{\citenamefont {Fisher}\ \emph {et~al.}(2020)\citenamefont {Fisher},
  \citenamefont {Sun},\ and\ \citenamefont {Kolemen}}]{FisherNF20}%
  \BibitemOpen
  \bibfield  {author} {\bibinfo {author} {\bibfnamefont {A.~E.}\ \bibnamefont
  {Fisher}}, \bibinfo {author} {\bibfnamefont {Z.}~\bibnamefont {Sun}}, \ and\
  \bibinfo {author} {\bibfnamefont {E.}~\bibnamefont {Kolemen}},\ }\href
  {\doibase 10.1016/j.nme.2020.100855} {\bibfield  {journal} {\bibinfo
  {journal} {Nuclear Materials and Energy}\ }\textbf {\bibinfo {volume} {25}},\
  \bibinfo {pages} {100855} (\bibinfo {year} {2020})}\BibitemShut {NoStop}%
\bibitem [{\citenamefont {Rindt}\ \emph {et~al.}(2021)\citenamefont {Rindt},
  \citenamefont {van~den Eijnden}, \citenamefont {Morgan},\ and\ \citenamefont
  {Cardozo}}]{RindtFED21}%
  \BibitemOpen
  \bibfield  {author} {\bibinfo {author} {\bibfnamefont {P.}~\bibnamefont
  {Rindt}}, \bibinfo {author} {\bibfnamefont {J.}~\bibnamefont {van~den
  Eijnden}}, \bibinfo {author} {\bibfnamefont {T.}~\bibnamefont {Morgan}}, \
  and\ \bibinfo {author} {\bibfnamefont {N.~L.}\ \bibnamefont {Cardozo}},\
  }\href {\doibase 10.1016/j.fusengdes.2021.112812} {\bibfield  {journal}
  {\bibinfo  {journal} {Fusion Engineering and Design}\ }\textbf {\bibinfo
  {volume} {173}},\ \bibinfo {pages} {112812} (\bibinfo {year}
  {2021})}\BibitemShut {NoStop}%
\bibitem [{\citenamefont {Suarez}\ \emph {et~al.}(2021)\citenamefont {Suarez},
  \citenamefont {Iraola}, \citenamefont {Lamp{\'o}n}, \citenamefont {de~les
  Valls},\ and\ \citenamefont {Batet}}]{SuarezFED21}%
  \BibitemOpen
  \bibfield  {author} {\bibinfo {author} {\bibfnamefont {D.}~\bibnamefont
  {Suarez}}, \bibinfo {author} {\bibfnamefont {E.}~\bibnamefont {Iraola}},
  \bibinfo {author} {\bibfnamefont {C.}~\bibnamefont {Lamp{\'o}n}}, \bibinfo
  {author} {\bibfnamefont {E.}~\bibnamefont {de~les Valls}}, \ and\ \bibinfo
  {author} {\bibfnamefont {L.}~\bibnamefont {Batet}},\ }\href {\doibase
  10.1016/j.fusengdes.2021.112503} {\bibfield  {journal} {\bibinfo  {journal}
  {Fusion Engineering and Design}\ }\textbf {\bibinfo {volume} {170}},\
  \bibinfo {pages} {112503} (\bibinfo {year} {2021})}\BibitemShut {NoStop}%
\bibitem [{\citenamefont {Saenz}\ \emph {et~al.}(2022)\citenamefont {Saenz},
  \citenamefont {Sun}, \citenamefont {Fisher}, \citenamefont {Wynne},\ and\
  \citenamefont {Kolemen}}]{SaenzNF22}%
  \BibitemOpen
  \bibfield  {author} {\bibinfo {author} {\bibfnamefont {F.}~\bibnamefont
  {Saenz}}, \bibinfo {author} {\bibfnamefont {Z.}~\bibnamefont {Sun}}, \bibinfo
  {author} {\bibfnamefont {A.~E.}\ \bibnamefont {Fisher}}, \bibinfo {author}
  {\bibfnamefont {B.}~\bibnamefont {Wynne}}, \ and\ \bibinfo {author}
  {\bibfnamefont {E.}~\bibnamefont {Kolemen}},\ }\href {\doibase
  10.1088/1741-4326/ac6682} {\bibfield  {journal} {\bibinfo  {journal} {Nuclear
  Fusion}\ }\textbf {\bibinfo {volume} {62}},\ \bibinfo {pages} {086008}
  (\bibinfo {year} {2022})}\BibitemShut {NoStop}%
\bibitem [{\citenamefont {Sergeev}\ \emph {et~al.}(2015)\citenamefont
  {Sergeev}, \citenamefont {Kuteev}, \citenamefont {Bykov}, \citenamefont
  {Gervash}, \citenamefont {Glazunov}, \citenamefont {Goncharov}, \citenamefont
  {Dnestrovskij}, \citenamefont {Khayrutdinov}, \citenamefont {Klishchenko},
  \citenamefont {Lukash}, \citenamefont {Mazul}, \citenamefont {Molchanov},
  \citenamefont {Petrov}, \citenamefont {Rozhansky}, \citenamefont {Shpanskiy},
  \citenamefont {Sivak}, \citenamefont {Skokov},\ and\ \citenamefont
  {Spitsyn}}]{SergeevNF15}%
  \BibitemOpen
  \bibfield  {author} {\bibinfo {author} {\bibfnamefont {V.}~\bibnamefont
  {Sergeev}}, \bibinfo {author} {\bibfnamefont {B.}~\bibnamefont {Kuteev}},
  \bibinfo {author} {\bibfnamefont {A.}~\bibnamefont {Bykov}}, \bibinfo
  {author} {\bibfnamefont {A.}~\bibnamefont {Gervash}}, \bibinfo {author}
  {\bibfnamefont {D.}~\bibnamefont {Glazunov}}, \bibinfo {author}
  {\bibfnamefont {P.}~\bibnamefont {Goncharov}}, \bibinfo {author}
  {\bibfnamefont {A.}~\bibnamefont {Dnestrovskij}}, \bibinfo {author}
  {\bibfnamefont {R.}~\bibnamefont {Khayrutdinov}}, \bibinfo {author}
  {\bibfnamefont {A.}~\bibnamefont {Klishchenko}}, \bibinfo {author}
  {\bibfnamefont {V.}~\bibnamefont {Lukash}}, \bibinfo {author} {\bibfnamefont
  {I.}~\bibnamefont {Mazul}}, \bibinfo {author} {\bibfnamefont
  {P.}~\bibnamefont {Molchanov}}, \bibinfo {author} {\bibfnamefont
  {V.}~\bibnamefont {Petrov}}, \bibinfo {author} {\bibfnamefont
  {V.}~\bibnamefont {Rozhansky}}, \bibinfo {author} {\bibfnamefont
  {Y.}~\bibnamefont {Shpanskiy}}, \bibinfo {author} {\bibfnamefont
  {A.}~\bibnamefont {Sivak}}, \bibinfo {author} {\bibfnamefont
  {V.}~\bibnamefont {Skokov}}, \ and\ \bibinfo {author} {\bibfnamefont
  {A.}~\bibnamefont {Spitsyn}},\ }\href {\doibase
  10.1088/0029-5515/55/12/123013} {\bibfield  {journal} {\bibinfo  {journal}
  {Nuclear Fusion}\ }\textbf {\bibinfo {volume} {55}},\ \bibinfo {pages}
  {123013} (\bibinfo {year} {2015})}\BibitemShut {NoStop}%
\bibitem [{\citenamefont {Pintsuk}\ \emph {et~al.}(2022)\citenamefont
  {Pintsuk}, \citenamefont {Aiello}, \citenamefont {Dudarev}, \citenamefont
  {Gorley}, \citenamefont {Henry}, \citenamefont {Richou}, \citenamefont
  {Rieth}, \citenamefont {Terentyev},\ and\ \citenamefont
  {Vila}}]{PintsukFED22}%
  \BibitemOpen
  \bibfield  {author} {\bibinfo {author} {\bibfnamefont {G.}~\bibnamefont
  {Pintsuk}}, \bibinfo {author} {\bibfnamefont {G.}~\bibnamefont {Aiello}},
  \bibinfo {author} {\bibfnamefont {S.}~\bibnamefont {Dudarev}}, \bibinfo
  {author} {\bibfnamefont {M.}~\bibnamefont {Gorley}}, \bibinfo {author}
  {\bibfnamefont {J.}~\bibnamefont {Henry}}, \bibinfo {author} {\bibfnamefont
  {M.}~\bibnamefont {Richou}}, \bibinfo {author} {\bibfnamefont
  {M.}~\bibnamefont {Rieth}}, \bibinfo {author} {\bibfnamefont
  {D.}~\bibnamefont {Terentyev}}, \ and\ \bibinfo {author} {\bibfnamefont
  {R.}~\bibnamefont {Vila}},\ }\href {\doibase 10.1016/j.fusengdes.2021.112994}
  {\bibfield  {journal} {\bibinfo  {journal} {Fusion Engineering and Design}\
  }\textbf {\bibinfo {volume} {174}},\ \bibinfo {pages} {112994} (\bibinfo
  {year} {2022})}\BibitemShut {NoStop}%
\bibitem [{\citenamefont {Zhang}\ \emph {et~al.}(2025)\citenamefont {Zhang},
  \citenamefont {Li},\ and\ \citenamefont {Liu}}]{ZhangAA25}%
  \BibitemOpen
  \bibfield  {author} {\bibinfo {author} {\bibfnamefont {J.}~\bibnamefont
  {Zhang}}, \bibinfo {author} {\bibfnamefont {J.}~\bibnamefont {Li}}, \ and\
  \bibinfo {author} {\bibfnamefont {X.}~\bibnamefont {Liu}},\ }\href {\doibase
  10.1063/5.0257060} {\bibfield  {journal} {\bibinfo  {journal} {AIP Advances}\
  }\textbf {\bibinfo {volume} {15}},\ \bibinfo {pages} {035333} (\bibinfo
  {year} {2025})}\BibitemShut {NoStop}%
\bibitem [{\citenamefont {Morley}\ \emph {et~al.}(2004)\citenamefont {Morley},
  \citenamefont {Smolentsev}, \citenamefont {Munipalli}, \citenamefont {Ni},
  \citenamefont {Gao},\ and\ \citenamefont {Abdou}}]{MorleyFED04}%
  \BibitemOpen
  \bibfield  {author} {\bibinfo {author} {\bibfnamefont {N.~B.}\ \bibnamefont
  {Morley}}, \bibinfo {author} {\bibfnamefont {S.}~\bibnamefont {Smolentsev}},
  \bibinfo {author} {\bibfnamefont {R.}~\bibnamefont {Munipalli}}, \bibinfo
  {author} {\bibfnamefont {M.-J.}\ \bibnamefont {Ni}}, \bibinfo {author}
  {\bibfnamefont {D.}~\bibnamefont {Gao}}, \ and\ \bibinfo {author}
  {\bibfnamefont {M.}~\bibnamefont {Abdou}},\ }\href {\doibase
  10.1016/j.fusengdes.2004.07.013} {\bibfield  {journal} {\bibinfo  {journal}
  {Fusion Engineering and Design}\ }\textbf {\bibinfo {volume} {72}},\ \bibinfo
  {pages} {3} (\bibinfo {year} {2004})}\BibitemShut {NoStop}%
\bibitem [{\citenamefont {Samulyak}\ \emph {et~al.}(2007)\citenamefont
  {Samulyak}, \citenamefont {Du}, \citenamefont {Glimm},\ and\ \citenamefont
  {Xu}}]{SamulyakJCP07}%
  \BibitemOpen
  \bibfield  {author} {\bibinfo {author} {\bibfnamefont {R.}~\bibnamefont
  {Samulyak}}, \bibinfo {author} {\bibfnamefont {J.}~\bibnamefont {Du}},
  \bibinfo {author} {\bibfnamefont {J.}~\bibnamefont {Glimm}}, \ and\ \bibinfo
  {author} {\bibfnamefont {Z.}~\bibnamefont {Xu}},\ }\href {\doibase
  10.1016/j.jcp.2007.06.005} {\bibfield  {journal} {\bibinfo  {journal}
  {Journal of Computational Physics}\ }\textbf {\bibinfo {volume} {226}},\
  \bibinfo {pages} {1532} (\bibinfo {year} {2007})}\BibitemShut {NoStop}%
\bibitem [{\citenamefont {Siriano}\ \emph {et~al.}(2024)\citenamefont
  {Siriano}, \citenamefont {Melchiorri}, \citenamefont {Pignatiello},\ and\
  \citenamefont {Tassone}}]{SirianoFED24}%
  \BibitemOpen
  \bibfield  {author} {\bibinfo {author} {\bibfnamefont {S.}~\bibnamefont
  {Siriano}}, \bibinfo {author} {\bibfnamefont {L.}~\bibnamefont {Melchiorri}},
  \bibinfo {author} {\bibfnamefont {S.}~\bibnamefont {Pignatiello}}, \ and\
  \bibinfo {author} {\bibfnamefont {A.}~\bibnamefont {Tassone}},\ }\href
  {\doibase 10.1016/j.fusengdes.2024.114216} {\bibfield  {journal} {\bibinfo
  {journal} {Fusion Engineering and Design}\ }\textbf {\bibinfo {volume}
  {200}},\ \bibinfo {pages} {114216} (\bibinfo {year} {2024})}\BibitemShut
  {NoStop}%
\bibitem [{\citenamefont {Al-Salami}(2022)}]{AlSalamiKU22}%
  \BibitemOpen
  \bibfield  {author} {\bibinfo {author} {\bibfnamefont {J.}~\bibnamefont
  {Al-Salami}},\ }\emph {\bibinfo {title} {Numerical simulation of MHD free
  surface liquid metal flows for nuclear fusion applications}},\ \href@noop {}
  {Ph.D. thesis},\ \bibinfo  {school} {Kyushu University} (\bibinfo {year}
  {2022})\BibitemShut {NoStop}%
\bibitem [{\citenamefont {Suarez}\ \emph {et~al.}(2022)\citenamefont {Suarez},
  \citenamefont {Khodak},\ and\ \citenamefont {de~les Valls}}]{SuarezITPS22}%
  \BibitemOpen
  \bibfield  {author} {\bibinfo {author} {\bibfnamefont {D.}~\bibnamefont
  {Suarez}}, \bibinfo {author} {\bibfnamefont {A.}~\bibnamefont {Khodak}}, \
  and\ \bibinfo {author} {\bibfnamefont {E.}~\bibnamefont {de~les Valls}},\
  }\href {\doibase 10.1109/TPS.2022.3203801} {\bibfield  {journal} {\bibinfo
  {journal} {IEEE Transactions on Plasma Science}\ }\textbf {\bibinfo {volume}
  {50}},\ \bibinfo {pages} {4206} (\bibinfo {year} {2022})}\BibitemShut
  {NoStop}%
\bibitem [{\citenamefont {Smolentsev}\ \emph {et~al.}(2010)\citenamefont
  {Smolentsev}, \citenamefont {Moreau}, \citenamefont {B{\"u}hler},\ and\
  \citenamefont {Mistrangelo}}]{SmolentsevFED10}%
  \BibitemOpen
  \bibfield  {author} {\bibinfo {author} {\bibfnamefont {S.}~\bibnamefont
  {Smolentsev}}, \bibinfo {author} {\bibfnamefont {R.}~\bibnamefont {Moreau}},
  \bibinfo {author} {\bibfnamefont {L.}~\bibnamefont {B{\"u}hler}}, \ and\
  \bibinfo {author} {\bibfnamefont {C.}~\bibnamefont {Mistrangelo}},\ }\href
  {\doibase 10.1016/j.fusengdes.2010.02.038} {\bibfield  {journal} {\bibinfo
  {journal} {Fusion Engineering and Design}\ }\textbf {\bibinfo {volume}
  {85}},\ \bibinfo {pages} {1196} (\bibinfo {year} {2010})}\BibitemShut
  {NoStop}%
\bibitem [{\citenamefont {Kawczynski}\ \emph {et~al.}(2016)\citenamefont
  {Kawczynski}, \citenamefont {Smolentsev},\ and\ \citenamefont
  {Abdou}}]{KawczynskiFED16}%
  \BibitemOpen
  \bibfield  {author} {\bibinfo {author} {\bibfnamefont {C.}~\bibnamefont
  {Kawczynski}}, \bibinfo {author} {\bibfnamefont {S.}~\bibnamefont
  {Smolentsev}}, \ and\ \bibinfo {author} {\bibfnamefont {M.}~\bibnamefont
  {Abdou}},\ }\href {\doibase 10.1016/j.fusengdes.2016.02.088} {\bibfield
  {journal} {\bibinfo  {journal} {Fusion Engineering and Design}\ }\textbf
  {\bibinfo {volume} {109-111}},\ \bibinfo {pages} {422} (\bibinfo {year}
  {2016})}\BibitemShut {NoStop}%
\bibitem [{\citenamefont {Kawczynski}\ \emph {et~al.}(2018)\citenamefont
  {Kawczynski}, \citenamefont {Smolentsev},\ and\ \citenamefont
  {Abdou}}]{KawczynskiPoF18}%
  \BibitemOpen
  \bibfield  {author} {\bibinfo {author} {\bibfnamefont {C.}~\bibnamefont
  {Kawczynski}}, \bibinfo {author} {\bibfnamefont {S.}~\bibnamefont
  {Smolentsev}}, \ and\ \bibinfo {author} {\bibfnamefont {M.}~\bibnamefont
  {Abdou}},\ }\href {\doibase 10.1063/1.5036775} {\bibfield  {journal}
  {\bibinfo  {journal} {Physics of Fluids}\ }\textbf {\bibinfo {volume} {30}},\
  \bibinfo {pages} {067103} (\bibinfo {year} {2018})}\BibitemShut {NoStop}%
\bibitem [{\citenamefont {Smolyanov}\ and\ \citenamefont
  {Zikanov}(2025)}]{SmolyanovPRE25}%
  \BibitemOpen
  \bibfield  {author} {\bibinfo {author} {\bibfnamefont {I.}~\bibnamefont
  {Smolyanov}}\ and\ \bibinfo {author} {\bibfnamefont {O.}~\bibnamefont
  {Zikanov}},\ }\href {\doibase 10.1103/nm74-f77r} {\bibfield  {journal}
  {\bibinfo  {journal} {Physical Review E}\ }\textbf {\bibinfo {volume}
  {111}},\ \bibinfo {pages} {065104} (\bibinfo {year} {2025})}\BibitemShut
  {NoStop}%
\bibitem [{\citenamefont {Levitt}\ \emph {et~al.}(2023)\citenamefont {Levitt},
  \citenamefont {Meier}, \citenamefont {Umstattd}, \citenamefont {Barhydt},
  \citenamefont {Datta}, \citenamefont {Liekhus-Schmaltz}, \citenamefont
  {Sutherland},\ and\ \citenamefont {Nelson}}]{LevittPoP23}%
  \BibitemOpen
  \bibfield  {author} {\bibinfo {author} {\bibfnamefont {B.}~\bibnamefont
  {Levitt}}, \bibinfo {author} {\bibfnamefont {E.~T.}\ \bibnamefont {Meier}},
  \bibinfo {author} {\bibfnamefont {R.}~\bibnamefont {Umstattd}}, \bibinfo
  {author} {\bibfnamefont {J.~R.}\ \bibnamefont {Barhydt}}, \bibinfo {author}
  {\bibfnamefont {I.~A.~M.}\ \bibnamefont {Datta}}, \bibinfo {author}
  {\bibfnamefont {C.}~\bibnamefont {Liekhus-Schmaltz}}, \bibinfo {author}
  {\bibfnamefont {D.~A.}\ \bibnamefont {Sutherland}}, \ and\ \bibinfo {author}
  {\bibfnamefont {B.~A.}\ \bibnamefont {Nelson}},\ }\href {\doibase
  10.1063/5.0163361} {\bibfield  {journal} {\bibinfo  {journal} {Physics of
  Plasmas}\ }\textbf {\bibinfo {volume} {30}},\ \bibinfo {pages} {090603}
  (\bibinfo {year} {2023})}\BibitemShut {NoStop}%
\bibitem [{\citenamefont {Endeve}\ \emph {et~al.}(2026)\citenamefont {Endeve},
  \citenamefont {Stefanski}, \citenamefont {Delchini}, \citenamefont
  {Slattery}, \citenamefont {Hauck}, \citenamefont {Turcksin},\ and\
  \citenamefont {Smolentsev}}]{EndeveFED26}%
  \BibitemOpen
  \bibfield  {author} {\bibinfo {author} {\bibfnamefont {E.}~\bibnamefont
  {Endeve}}, \bibinfo {author} {\bibfnamefont {D.}~\bibnamefont {Stefanski}},
  \bibinfo {author} {\bibfnamefont {M.-O.~G.}\ \bibnamefont {Delchini}},
  \bibinfo {author} {\bibfnamefont {S.}~\bibnamefont {Slattery}}, \bibinfo
  {author} {\bibfnamefont {C.~D.}\ \bibnamefont {Hauck}}, \bibinfo {author}
  {\bibfnamefont {B.}~\bibnamefont {Turcksin}}, \ and\ \bibinfo {author}
  {\bibfnamefont {S.}~\bibnamefont {Smolentsev}},\ }\href {\doibase
  10.1016/j.fusengdes.2026.115792} {\bibfield  {journal} {\bibinfo  {journal}
  {Fusion Engineering and Design}\ }\textbf {\bibinfo {volume} {228}},\
  \bibinfo {pages} {115792} (\bibinfo {year} {2026})}\BibitemShut {NoStop}%
\bibitem [{\citenamefont {Blishchik}\ \emph {et~al.}(2021)\citenamefont
  {Blishchik}, \citenamefont {van~der Lans},\ and\ \citenamefont {Kenjere{\v
  s}}}]{BlishchikIJHFF21}%
  \BibitemOpen
  \bibfield  {author} {\bibinfo {author} {\bibfnamefont {A.}~\bibnamefont
  {Blishchik}}, \bibinfo {author} {\bibfnamefont {M.}~\bibnamefont {van~der
  Lans}}, \ and\ \bibinfo {author} {\bibfnamefont {S.}~\bibnamefont {Kenjere{\v
  s}}},\ }\href {\doibase 10.1016/j.ijheatfluidflow.2021.108800} {\bibfield
  {journal} {\bibinfo  {journal} {International Journal of Heat and Fluid
  Flow}\ }\textbf {\bibinfo {volume} {90}},\ \bibinfo {pages} {108800}
  (\bibinfo {year} {2021})}\BibitemShut {NoStop}%
\bibitem [{\citenamefont {Suarez}\ and\ \citenamefont
  {Smolentsev}(2025)}]{SuarezFED25}%
  \BibitemOpen
  \bibfield  {author} {\bibinfo {author} {\bibfnamefont {D.}~\bibnamefont
  {Suarez}}\ and\ \bibinfo {author} {\bibfnamefont {S.}~\bibnamefont
  {Smolentsev}},\ }\href {\doibase 10.1016/j.fusengdes.2025.115008} {\bibfield
  {journal} {\bibinfo  {journal} {Fusion Engineering and Design}\ }\textbf
  {\bibinfo {volume} {215}},\ \bibinfo {pages} {115008} (\bibinfo {year}
  {2025})}\BibitemShut {NoStop}%
\bibitem [{\citenamefont {Suponitsky}\ \emph {et~al.}(2022)\citenamefont
  {Suponitsky}, \citenamefont {Khalzov},\ and\ \citenamefont
  {Avital}}]{SuponitskyFluids22}%
  \BibitemOpen
  \bibfield  {author} {\bibinfo {author} {\bibfnamefont {V.}~\bibnamefont
  {Suponitsky}}, \bibinfo {author} {\bibfnamefont {I.~V.}\ \bibnamefont
  {Khalzov}}, \ and\ \bibinfo {author} {\bibfnamefont {E.~J.}\ \bibnamefont
  {Avital}},\ }\href {\doibase 10.3390/fluids7070210} {\bibfield  {journal}
  {\bibinfo  {journal} {Fluids}\ }\textbf {\bibinfo {volume} {7}},\ \bibinfo
  {pages} {210} (\bibinfo {year} {2022})}\BibitemShut {NoStop}%
\bibitem [{\citenamefont {Suponitsky}\ \emph {et~al.}(2025)\citenamefont
  {Suponitsky}, \citenamefont {Khalzov}, \citenamefont {Roberts},\ and\
  \citenamefont {Forysinski}}]{SuponitskyFluids25}%
  \BibitemOpen
  \bibfield  {author} {\bibinfo {author} {\bibfnamefont {V.}~\bibnamefont
  {Suponitsky}}, \bibinfo {author} {\bibfnamefont {I.~V.}\ \bibnamefont
  {Khalzov}}, \bibinfo {author} {\bibfnamefont {D.~M.}\ \bibnamefont
  {Roberts}}, \ and\ \bibinfo {author} {\bibfnamefont {P.~W.}\ \bibnamefont
  {Forysinski}},\ }\href {\doibase 10.3390/fluids10090222} {\bibfield
  {journal} {\bibinfo  {journal} {Fluids}\ }\textbf {\bibinfo {volume} {10}},\
  \bibinfo {pages} {222} (\bibinfo {year} {2025})}\BibitemShut {NoStop}%
\bibitem [{\citenamefont {Wynne}\ \emph {et~al.}(2025)\citenamefont {Wynne},
  \citenamefont {Saenz}, \citenamefont {Al-Salami}, \citenamefont {Xu},
  \citenamefont {Sun}, \citenamefont {Hu}, \citenamefont {Hanada},\ and\
  \citenamefont {Kolemen}}]{WynnePoP25}%
  \BibitemOpen
  \bibfield  {author} {\bibinfo {author} {\bibfnamefont {B.}~\bibnamefont
  {Wynne}}, \bibinfo {author} {\bibfnamefont {F.}~\bibnamefont {Saenz}},
  \bibinfo {author} {\bibfnamefont {J.}~\bibnamefont {Al-Salami}}, \bibinfo
  {author} {\bibfnamefont {Y.}~\bibnamefont {Xu}}, \bibinfo {author}
  {\bibfnamefont {Z.}~\bibnamefont {Sun}}, \bibinfo {author} {\bibfnamefont
  {C.}~\bibnamefont {Hu}}, \bibinfo {author} {\bibfnamefont {K.}~\bibnamefont
  {Hanada}}, \ and\ \bibinfo {author} {\bibfnamefont {E.}~\bibnamefont
  {Kolemen}},\ }\href {\doibase 10.1063/5.0230242} {\bibfield  {journal}
  {\bibinfo  {journal} {Physics of Plasmas}\ }\textbf {\bibinfo {volume}
  {32}},\ \bibinfo {pages} {013907} (\bibinfo {year} {2025})}\BibitemShut
  {NoStop}%
\bibitem [{\citenamefont {Weller}\ \emph {et~al.}(1998)\citenamefont {Weller},
  \citenamefont {Tabor}, \citenamefont {Jasak},\ and\ \citenamefont
  {Fureby}}]{WellerCP98}%
  \BibitemOpen
  \bibfield  {author} {\bibinfo {author} {\bibfnamefont {H.~G.}\ \bibnamefont
  {Weller}}, \bibinfo {author} {\bibfnamefont {G.}~\bibnamefont {Tabor}},
  \bibinfo {author} {\bibfnamefont {H.}~\bibnamefont {Jasak}}, \ and\ \bibinfo
  {author} {\bibfnamefont {C.}~\bibnamefont {Fureby}},\ }\href {\doibase
  10.1063/1.168744} {\bibfield  {journal} {\bibinfo  {journal} {Computers in
  Physics}\ }\textbf {\bibinfo {volume} {12}},\ \bibinfo {pages} {620}
  (\bibinfo {year} {1998})}\BibitemShut {NoStop}%
\bibitem [{\citenamefont {Issa}(1986)}]{IssaJCP86}%
  \BibitemOpen
  \bibfield  {author} {\bibinfo {author} {\bibfnamefont {R.}~\bibnamefont
  {Issa}},\ }\href {\doibase 10.1016/0021-9991(86)90099-9} {\bibfield
  {journal} {\bibinfo  {journal} {Journal of Computational Physics}\ }\textbf
  {\bibinfo {volume} {62}},\ \bibinfo {pages} {40} (\bibinfo {year}
  {1986})}\BibitemShut {NoStop}%
\bibitem [{\citenamefont {Caretto}\ \emph {et~al.}(1973)\citenamefont
  {Caretto}, \citenamefont {Gosman}, \citenamefont {Patankar},\ and\
  \citenamefont {Spalding}}]{CarettoProc73}%
  \BibitemOpen
  \bibfield  {author} {\bibinfo {author} {\bibfnamefont {L.}~\bibnamefont
  {Caretto}}, \bibinfo {author} {\bibfnamefont {A.}~\bibnamefont {Gosman}},
  \bibinfo {author} {\bibfnamefont {S.}~\bibnamefont {Patankar}}, \ and\
  \bibinfo {author} {\bibfnamefont {D.}~\bibnamefont {Spalding}},\ }in\ \href
  {\doibase 10.1007/BFb0112677} {\emph {\bibinfo {booktitle} {Proceedings of
  the Third International Conference on Numerical Methods in Fluid
  Mechanics}}},\ \bibinfo {editor} {edited by\ \bibinfo {editor} {\bibfnamefont
  {H.}~\bibnamefont {Cabannes}}\ and\ \bibinfo {editor} {\bibfnamefont
  {R.}~\bibnamefont {Temam}}}\ (\bibinfo  {publisher} {Springer Berlin
  Heidelberg},\ \bibinfo {year} {1973})\ pp.\ \bibinfo {pages}
  {60--68}\BibitemShut {NoStop}%
\bibitem [{\citenamefont {Deshpande}\ \emph {et~al.}(2012)\citenamefont
  {Deshpande}, \citenamefont {Anumolu},\ and\ \citenamefont
  {Trujillo}}]{DeshpandeCSD12}%
  \BibitemOpen
  \bibfield  {author} {\bibinfo {author} {\bibfnamefont {S.}~\bibnamefont
  {Deshpande}}, \bibinfo {author} {\bibfnamefont {L.}~\bibnamefont {Anumolu}},
  \ and\ \bibinfo {author} {\bibfnamefont {M.}~\bibnamefont {Trujillo}},\
  }\href {\doibase 10.1088/1749-4699/5/1/014016} {\bibfield  {journal}
  {\bibinfo  {journal} {Computational Science \& Discovery}\ }\textbf {\bibinfo
  {volume} {5}},\ \bibinfo {pages} {014016} (\bibinfo {year}
  {2012})}\BibitemShut {NoStop}%
\bibitem [{\citenamefont {Brackbill}\ \emph {et~al.}(1992)\citenamefont
  {Brackbill}, \citenamefont {Kothe},\ and\ \citenamefont
  {Zemach}}]{BrackbillJCP92}%
  \BibitemOpen
  \bibfield  {author} {\bibinfo {author} {\bibfnamefont {J.}~\bibnamefont
  {Brackbill}}, \bibinfo {author} {\bibfnamefont {D.}~\bibnamefont {Kothe}}, \
  and\ \bibinfo {author} {\bibfnamefont {C.}~\bibnamefont {Zemach}},\ }\href
  {\doibase 10.1016/0021-9991(92)90240-Y} {\bibfield  {journal} {\bibinfo
  {journal} {Journal of Computational Physics}\ }\textbf {\bibinfo {volume}
  {100}},\ \bibinfo {pages} {335} (\bibinfo {year} {1992})}\BibitemShut
  {NoStop}%
\bibitem [{\citenamefont {Ni}\ \emph {et~al.}(2007{\natexlab{a}})\citenamefont
  {Ni}, \citenamefont {Munipalli}, \citenamefont {Morley}, \citenamefont
  {Huang},\ and\ \citenamefont {Abdou}}]{NiJCP07a}%
  \BibitemOpen
  \bibfield  {author} {\bibinfo {author} {\bibfnamefont {M.-J.}\ \bibnamefont
  {Ni}}, \bibinfo {author} {\bibfnamefont {R.}~\bibnamefont {Munipalli}},
  \bibinfo {author} {\bibfnamefont {N.~B.}\ \bibnamefont {Morley}}, \bibinfo
  {author} {\bibfnamefont {P.}~\bibnamefont {Huang}}, \ and\ \bibinfo {author}
  {\bibfnamefont {M.~A.}\ \bibnamefont {Abdou}},\ }\href {\doibase
  10.1016/j.jcp.2007.07.025} {\bibfield  {journal} {\bibinfo  {journal}
  {Journal of Computational Physics}\ }\textbf {\bibinfo {volume} {227}},\
  \bibinfo {pages} {174} (\bibinfo {year} {2007}{\natexlab{a}})}\BibitemShut
  {NoStop}%
\bibitem [{\citenamefont {Ni}\ \emph {et~al.}(2007{\natexlab{b}})\citenamefont
  {Ni}, \citenamefont {Munipalli}, \citenamefont {Huang}, \citenamefont
  {Morley},\ and\ \citenamefont {Abdou}}]{NiJCP07b}%
  \BibitemOpen
  \bibfield  {author} {\bibinfo {author} {\bibfnamefont {M.-J.}\ \bibnamefont
  {Ni}}, \bibinfo {author} {\bibfnamefont {R.}~\bibnamefont {Munipalli}},
  \bibinfo {author} {\bibfnamefont {P.}~\bibnamefont {Huang}}, \bibinfo
  {author} {\bibfnamefont {N.~B.}\ \bibnamefont {Morley}}, \ and\ \bibinfo
  {author} {\bibfnamefont {M.~A.}\ \bibnamefont {Abdou}},\ }\href {\doibase
  10.1016/j.jcp.2007.07.023} {\bibfield  {journal} {\bibinfo  {journal}
  {Journal of Computational Physics}\ }\textbf {\bibinfo {volume} {227}},\
  \bibinfo {pages} {205} (\bibinfo {year} {2007}{\natexlab{b}})}\BibitemShut
  {NoStop}%
\bibitem [{\citenamefont {M{\"u}ller}\ and\ \citenamefont
  {B{\"u}hler}(2001)}]{MullerBook}%
  \BibitemOpen
  \bibfield  {author} {\bibinfo {author} {\bibfnamefont {U.}~\bibnamefont
  {M{\"u}ller}}\ and\ \bibinfo {author} {\bibfnamefont {L.}~\bibnamefont
  {B{\"u}hler}},\ }\href {\doibase 10.1007/978-3-662-04405-6} {\emph {\bibinfo
  {title} {Magnetofluiddynamics in Channels and Containers}}}\ (\bibinfo
  {publisher} {Springer Berlin, Heidelberg},\ \bibinfo {year}
  {2001})\BibitemShut {NoStop}%
\bibitem [{\citenamefont {Smolentsev}(2025)}]{SmolentsevFED25}%
  \BibitemOpen
  \bibfield  {author} {\bibinfo {author} {\bibfnamefont {S.}~\bibnamefont
  {Smolentsev}},\ }\href {\doibase 10.1016/j.fusengdes.2025.114854} {\bibfield
  {journal} {\bibinfo  {journal} {Fusion Engineering and Design}\ }\textbf
  {\bibinfo {volume} {213}},\ \bibinfo {pages} {114854} (\bibinfo {year}
  {2025})}\BibitemShut {NoStop}%
\bibitem [{\citenamefont {Sun}\ \emph {et~al.}(2023)\citenamefont {Sun},
  \citenamefont {Al-Salami}, \citenamefont {Khodak}, \citenamefont {Saenz},
  \citenamefont {Wynne}, \citenamefont {Maingi}, \citenamefont {Hanada},
  \citenamefont {Hu},\ and\ \citenamefont {Kolemen}}]{SunNF23}%
  \BibitemOpen
  \bibfield  {author} {\bibinfo {author} {\bibfnamefont {Z.}~\bibnamefont
  {Sun}}, \bibinfo {author} {\bibfnamefont {J.}~\bibnamefont {Al-Salami}},
  \bibinfo {author} {\bibfnamefont {A.}~\bibnamefont {Khodak}}, \bibinfo
  {author} {\bibfnamefont {F.}~\bibnamefont {Saenz}}, \bibinfo {author}
  {\bibfnamefont {B.}~\bibnamefont {Wynne}}, \bibinfo {author} {\bibfnamefont
  {R.}~\bibnamefont {Maingi}}, \bibinfo {author} {\bibfnamefont
  {K.}~\bibnamefont {Hanada}}, \bibinfo {author} {\bibfnamefont
  {C.}~\bibnamefont {Hu}}, \ and\ \bibinfo {author} {\bibfnamefont
  {E.}~\bibnamefont {Kolemen}},\ }\href {\doibase 10.1088/1741-4326/acd864}
  {\bibfield  {journal} {\bibinfo  {journal} {Nuclear Fusion}\ }\textbf
  {\bibinfo {volume} {63}},\ \bibinfo {pages} {076022} (\bibinfo {year}
  {2023})}\BibitemShut {NoStop}%
\bibitem [{\citenamefont {Morley}\ \emph {et~al.}(2008)\citenamefont {Morley},
  \citenamefont {Burris}, \citenamefont {Cadwallader},\ and\ \citenamefont
  {Nornberg}}]{MorleyRSI08}%
  \BibitemOpen
  \bibfield  {author} {\bibinfo {author} {\bibfnamefont {N.}~\bibnamefont
  {Morley}}, \bibinfo {author} {\bibfnamefont {J.}~\bibnamefont {Burris}},
  \bibinfo {author} {\bibfnamefont {L.}~\bibnamefont {Cadwallader}}, \ and\
  \bibinfo {author} {\bibfnamefont {M.}~\bibnamefont {Nornberg}},\ }\href
  {\doibase 10.1063/1.2930813} {\bibfield  {journal} {\bibinfo  {journal}
  {Review of Scientific Instruments}\ }\textbf {\bibinfo {volume} {79}},\
  \bibinfo {pages} {056107} (\bibinfo {year} {2008})}\BibitemShut {NoStop}%
\bibitem [{https://github.com/PlasmaControl/FreeMHD()}]{FreeMHD}%
  \BibitemOpen
  https://github.com/PlasmaControl/FreeMHD,\ \href@noop {} {} (\bibinfo {year}
  {2026})\BibitemShut {NoStop}%
\bibitem [{\citenamefont {Jung}(2026)}]{Zenodo}%
  \BibitemOpen
  \bibfield  {author} {\bibinfo {author} {\bibfnamefont {M.~K.}\ \bibnamefont
  {Jung}},\ }\href {\doibase 10.5281/zenodo.20727181} {\bibfield  {journal}
  {\bibinfo  {journal} {Zenodo}\ } (\bibinfo {year} {2026}),\
  10.5281/zenodo.20727181}\BibitemShut {NoStop}%
\end{thebibliography}%

\end{document}